\documentclass[pra,twocolumn,superscriptaddress]{revtex4-1}

\usepackage[utf8]{inputenc} 
\usepackage{color}
\usepackage{bm}
\usepackage{amssymb,amsmath}
\usepackage{bbold}
\usepackage{hyperref}
\usepackage{xfrac}
\usepackage{makecell}
\usepackage[normalem]{ulem}

\newcommand{\ah}[0]{\hat{a}}
\newcommand{\dhat}[0]{\hat{d}}
\newcommand{\lcrit}[0]{\lambda_{\mathrm{crit}}}

\newcommand{\lsp}[0]{\lambda_{\mathrm{1}}}
\newcommand{\Dsp}[0]{\Delta_{\mathrm{1}}}

\newcommand{\corrSinglePump}[0]{\hat{H}_{\mathrm{1c}}}

\newcommand{\Ddp}[0]{\Delta_{\mathrm{2}}}
\newcommand{\Dtdp}[0]{\Delta_{\mathrm{2}}}
\newcommand{\ldp}[0]{\lambda_{\mathrm{2}}}

\newcommand{\corrDoublePump}[0]{\hat{H}_{\mathrm{2c}}}

\newcommand{\lfp}[0]{\lambda_{\mathrm{f}}}
\newcommand{\Dfp}[0]{\Delta_{\mathrm{f}}}
\newcommand{\Lfp}[0]{\Lambda_{\mathrm{f}}}

\newcommand{\corrFluxPump}[0]{\hat{H}_{\mathrm{fc}}}

\newcommand{\wpump}[0]{\omega_{\mathrm{p}}}
\newcommand{\df}[0]{\delta\!f}
\newcommand{\kappaTot}[0]{\overline{\kappa}}
\newcommand{\ffpz}[0]{\Phi_{\mathrm{zpf}}}

\newcommand{\ain}[0]{\hat{a}_{\mathrm{in}}}
\newcommand{\bin}[0]{\hat{b}_{\mathrm{in}}}

\newcommand{\aout}[0]{\hat{a}_{\mathrm{out}}}

\newcommand{\dout}[0]{\hat{d}_{\mathrm{out}}}

\newcommand{\parS}[1]{\left[#1\right]}
\newcommand{\parO}[1]{\left(#1\right)}
\newcommand{\parC}[1]{\left\{#1\right\}}
\newcommand{\matO}[1]{\parO{\begin{matrix}#1\end{matrix}}}
\newcommand{\matS}[1]{\parS{\begin{matrix}#1\end{matrix}}}

\newcommand{\abs}[1]{\left|#1\right|}

\newcommand{\comm}[2]{\left[#1 \textrm{,} #2 \right]}

\newcommand{\ket}[1]{|{#1}\rangle}

\newcommand{\average}[1]{\left\langle #1\right\rangle}
\newcommand{\cumulant}[1]{\langle \! \langle #1\rangle\! \rangle}
\newcommand{\drm}[0]{\mathrm{d}}
\newcommand{\erm}[0]{\mathrm{e}}
\newcommand{\Grm}[0]{\mathrm{g}}

\newcommand{\Am}[0]{\mathcal{A}}

\newcommand{\Omh}[0]{\hat{\mathcal{O}}}

\newcommand{\re}[1]{\mathrm{Re}\parC{#1}}

\begin{document}

\title{Effect of higher-order nonlinearities on amplification and squeezing in Josephson parametric amplifiers}

\author{Samuel Boutin}
\email{Samuel.Boutin@USherbrooke.ca}
\affiliation{Institut quantique et D\'{e}partement de Physique, Universit\'{e} de Sherbrooke, Sherbrooke, Qu\'{e}bec J1K 2R1, Canada}
\author{David M. Toyli} 
\author{ Aditya V. Venkatramani }
\altaffiliation{Current address: Department of Physics, Harvard University, Cambridge, Massachusetts, 02138, USA}
\author{  Andrew W. Eddins }
\author{  Irfan Siddiqi }
\affiliation{
Quantum Nanoelectronics Laboratory, Department of Physics, University of California, Berkeley, CA 94720, USA}
\affiliation{Center for Quantum Coherent Science, University of California, Berkeley CA 94720, USA}
\author{Alexandre Blais}
\affiliation{Institut quantique et D\'{e}partement de Physique, Universit\'{e} de Sherbrooke, Sherbrooke, Qu\'{e}bec J1K 2R1, Canada}
\affiliation{Canadian Institute for Advanced Research, Toronto, Canada}

\date{\today}                                       
\begin{abstract}
Single-mode Josephson junction-based parametric amplifiers are often modeled as perfect amplifiers and squeezers. We show that, in practice, the gain, quantum efficiency, and output field squeezing of these devices are limited by usually neglected higher-order corrections to the idealized model. To arrive at this result, we derive the leading corrections to the lumped-element Josephson parametric amplifier of three common pumping schemes: monochromatic current pump, bichromatic current pump, and monochromatic flux pump. We show that the leading correction for the last two schemes is a single Kerr-type quartic term, while the first scheme contains additional cubic terms. In all cases, we find that the corrections are detrimental to squeezing. In addition, we show that the Kerr correction leads to a strongly phase-dependent reduction of the quantum efficiency of a phase-sensitive measurement. Finally, we quantify the departure from ideal Gaussian character of the filtered output field from numerical calculation of third and fourth order cumulants. Our results show that, while a Gaussian output field is expected for an ideal Josephson parametric amplifier, higher-order corrections lead to non-Gaussian effects which increase with both gain and nonlinearity strength.
This theoretical study is complemented by experimental characterization of the output field of a flux-driven Josephson parametric amplifier. 
In addition to a measurement of the squeezing level of the filtered output field, the Husimi $Q$-function of the output field is imaged by the use of a deconvolution technique and compared to numerical results. 
This work establishes nonlinear corrections to the standard degenerate parametric amplifier model as an important contribution to Josephson parametric amplifier's squeezing and noise performance.
\end{abstract}

\maketitle

\section{Introduction}
Driven by the need for fast, high-fidelity single-shot readout of superconducting qubits,  superconducting low-noise microwave amplifiers are the subject of intense research.
Following the path of Yurke's \textit{et al.} work in the late 1980s~\cite{Yurke:1988kb,Yurke:1989fk,Yurke:1989kx}, several designs of Josephson junction-based parametric amplifiers (JPAs) have been introduced~\cite{Castellanos-Beltran:2007ys,Yamamoto:2008dp,Bergeal:2010uq,Hatridge:2011qf,Mutus:2013yu,Eichler:2013fk,Mutus:2014ve,Eichler:2014uq}. 
In addition to high-fidelity superconducting qubit readout leading to the observation of quantum jumps~\cite{Jeffrey:2014uq,Vijay:2011ve},
this new generation of near quantum-limited amplifiers have opened up new experimental possibilities such as
the creation and tomography of squeezed microwave light~\cite{Castellanos-Beltran:2008vn,Mallet:2011fk,Fedorov:2016}, and detailed weak measurement experiments~\cite{Murch:2013uq,Weber:2014fv,Campagne-Ibarcq:2014qf}.
JPAs are now ubiquitous in current superconducting circuit experiments, and applications in other research communities are growing~\cite{Stehlik:2015nr,Virally:2015fj,Simoneau2016,Goetz:2017}.

Depending on their design and operating mode, JPAs can fall in either of two broad categories of linear amplifiers: phase-preserving and phase-sensitive~\cite{caves:1982a,Roy:2016fk}. JPAs in the former category amplify both quadratures of the signal and quantum mechanics put a strict lower bound on the noise added by this process. On the contrary, JPAs in the latter category can amplify the signal of a single quadrature without any added noise by proportionally attenuating the conjugate quadrature. In other words, a phase-sensitive amplification is a source of squeezed radiation~\cite{Walls:2008fk,Gardiner:2004fk}. The properties of JPAs as a source of squeezed light are therefore intimately related to their noise properties as a phase-sensitive amplifier. While JPAs are usually modeled as quantum-limited amplifiers, and thus perfect squeezers, experimental results indicate that nonidealities limits both the achievable level of squeezing~\cite{Zhong:2013vn,Murch:2013kx,Bienfait:2016yq} and the measurement quantum efficiency~\cite{Murch:2013uq,Murch:2013kx,Vijay:2012uq,weberThesis}.

We show that these nonidealities are linked to higher-order corrections to the JPA Hamiltonian due to the Josephson cosine potential. We go beyond the standard analytical results by considering numerical solutions to the quantum master equation, including these usually neglected higher-order corrections. We derive the corrections to the JPA Hamiltonian for the single-mode and single-port lumped-element JPA~\cite{Roy:2016fk,Hatridge:2011qf}.
Using the formalism of quantum optics, we compare three frequently used pumping schemes of the JPA: monochromatic current pump~\cite{Hatridge:2011qf,Yurke:2006fk}, bichromatic current pump~\cite{Kamal:2009uq}, and monochromatic flux pump~\cite{Yamamoto:2008dp,Wustmann:2013uq,Zhou:2014fk}. We derive the leading higher-order corrections for each and study numerically their effect on gain, quantum efficiency, squeezing level and gaussianity of the output field. In addition, by comparing numerical results to an experimental characterization of the JPA output field, we show that the squeezing level saturation previously reported in the literature~\cite{Murch:2013kx,Zhong:2013vn} can be explained by including leading non-quadratic corrections in the Hamiltonian. The focus of this work on higher-order corrections complements recent theoretical investigations of various JPA designs~\cite{Yurke:2006fk,Kamal:2009uq,Wustmann:2013uq,Eichler:2013kx,Kochetov:2015ab,Roy:2016fk}.

The paper is organized as follows.
In Sec.~\ref{sec::DPA}, we set the notations and recall useful results for the standard quantum optics model of the degenerate parametric amplifier (DPA).
In Sec.~\ref{sec::LJPApumps}, we introduce the lumped-element JPA and the three different pumping schemes investigated in this work. For each scheme, we derive the higher-order corrections and show how, in the low gain and low nonlinearity regime, the system can be mapped back to the DPA. 
The respective advantages of the three pumping schemes are compared. In Sec.~\ref{sec::intracavitySignatures}, we compare the intracavity field properties of the different schemes including deviations from the results of the DPA. We also show that higher-order corrections can lead to non-Gaussian intracavity fields. In Sec.~\ref{sec::gainEta}, we characterize the JPA as an amplifier by calculating the gain and the quantum efficiency for both the phase-sensitive and phase-preserving modes of operation.
In order to relate these results to a series of experiments~\cite{Zhong:2013vn,Murch:2013kx,Bienfait:2016yq}, in Sec.~\ref{sec::filteredOutput}
we focus on the phase-sensitive amplification of vacuum, i.e.~squeezed vacuum. We characterize the JPA as a source of squeezed light by calculating moments and cumulants of the filtered output field. This allows us to obtain the squeezing level of the light generated, as well as to estimate its non-Gaussian character. In this section, numerical results are discussed together with an experimental characterization of the output field, including a direct imaging of the non-Gaussian distortions of the field due to nonidealities.
Finally, Sec.~\ref{sec::conclusion} summarizes our work.

\section{Degenerate parametric amplifier in a nutshell}
\label{sec::DPA}

To set the notation, we begin by presenting the DPA model and its solution~\cite{Gardiner:2004fk,Yurke:2006fk,Laflamme:2011vn}. In the next sections, we show how the JPA can be mapped to the DPA and study deviations from this simple model due to higher-order corrections. 
 
\begin{figure}[tb]
	\includegraphics[width=0.95\linewidth]{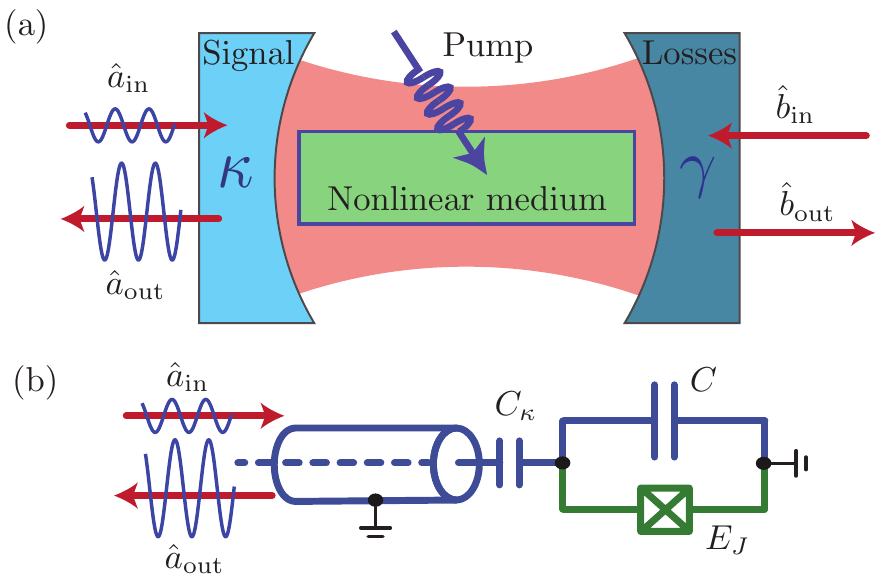}
	\caption{(a) Schematic of a DPA as a two-sided cavity in the optical domain. 
	(b) Circuit of a lossless lumped-element JPA, a reflection amplifier in the microwave domain.
	}
	\label{fig:circuit}
\end{figure}

The DPA is a standard model of quantum optics exhibiting parametric amplification and squeezing. 
In this model, as illustrated in Fig.~\ref{fig:circuit}(a), a nonlinear medium inserted in a single-mode cavity is pumped in order to modulate its refractive index at twice the cavity frequency~\cite{Walls:2008fk}. 
This modulates the effective length of the cavity and, as a consequence, its frequency. 
This modulation acts as an external source of energy leading to parametric amplification~\cite{Landau:1969uq,Nation:2012fk}. 

Introducing the annihilation (creation) operator $\ah^{(\dag)}$ for excitations in the cavity, the DPA Hamiltonian is (setting $\hbar=1$ for the remainder of the paper)
\begin{equation}
	H = \omega_c \ah^\dag  \ah  + \chi \parO{\ah^{\dag 2} \ah_p + \ah^2 \ah^\dag_p},
\end{equation}
with $\omega_c$ the cavity frequency, $\chi$ the nonlinearity, and $\ah_p^{(\dag)}$, the annihilation (creation) operator of an excitation in the external pump mode.
In the strong classical pump regime where $\ah_p \approx \alpha_p \erm^{-i \omega_p t}$ and in a frame rotating at half the parametric pump frequency $\wpump \sim 2 \omega_c$, the Hamiltonian is~\cite{Collett:1984kl,Walls:2008fk} 
\begin{equation}
	\hat{H}_{\mathrm{DPA}} = \Delta \ah^\dagger \ah  + \frac{\lambda}{2} \ah^{\dagger 2} + \frac{\lambda^*}{2} \ah^2,
\end{equation}
with the detuning $\Delta = \omega_{\mathrm{c}}- \wpump/2$ and $\lambda = 2 \chi \alpha_p$ the amplitude of the parametric pump.

Using input-output theory~\cite{Gardiner:2004fk,Walls:2008fk}, the equation of motion for the intracavity field is
\begin{equation} \label{eq::EOMDPA1}
	\dot \ah = i \comm{\hat{H}_{\mathrm{DPA}}}{\ah} - \frac{\kappaTot }{2}\ah +\sqrt{\kappa} \ain + \sqrt{\gamma} \bin,
\end{equation}
where the input mode $\ain$ (coupled to the cavity with rate $\kappa$) carries the signal to be amplified, while the input mode $\bin$ (coupled to the cavity with rate $\gamma$) mixes additional vacuum noise to the signal due to undesired losses. The total damping rate of the cavity is given by $\kappaTot = \kappa + \gamma$.

As shown in Appendix~\ref{app::solvingDPA}, the solution to Eq.~\eqref{eq::EOMDPA1} can be obtained in Fourier space. Using the boundary condition~\cite{Gardiner:2004fk,Walls:2008fk}
\begin{equation}
	\aout = \sqrt{\kappa} \ah - \ain,
	\label{eq:boundary}
\end{equation}
where $\aout$ is the output field carrying the amplified signal, 
one obtains the solution 
\begin{equation}
\begin{split}
	\overline{\aout} [\omega] &= 
	g_{S,\omega} \overline{\ain}[\omega] 
	+ 
	g_{I,\omega}\overline{\ain}^\dagger[-\omega]
	\\ &
	+
	\sqrt{\frac{\gamma}{\kappa}}
	\parS{
		\parO{g_{S,\omega}+1} \overline{\bin}[\omega] 
		+ 
		g_{I,\omega}\overline{\bin}^\dagger[-\omega]
	},
	\end{split}
	\label{eq::DPA_inputOutput}
\end{equation}
where the signal and idler amplitude gains are defined as
\begin{align}
	g_{S,\omega} &= \parS{ \frac{\kappa \kappaTot /2-i\kappa \parO{ \Delta+ \omega } }{D[\omega]} -1}, 
	\label{eq:gainSignal}
		\\ 
		g_{I,\omega} &= \frac{- i \kappa \lambda}{D[\omega]}, \label{eq:gainIdler}
\end{align}
with $D[\omega]= \Delta^2 + \parO{\kappaTot/2 - i \omega}^2- \abs{\lambda}^2$, and the frequency $\omega$ defined in the rotating frame such that a signal at $\omega=0$ is in resonance with the rotating frame frequency $\wpump/2$.
The output signal at frequency $\omega$ mixes and amplifies the input signal and idler modes at frequencies $\pm \omega$. 
In the lossless case ($\gamma=0$), the signal and idler gains obey $\abs{g_{S,\omega}}^2 = \abs{g_{I,\omega}}^2 +1$ and the input-output relation is a unitary squeezing transformation~\cite{caves:1982a}. On the contrary, in the presence of losses ($\gamma \neq 0$), additional noise is mixed with the input signal and the DPA is not a quantum-limited amplifier.

If the measurement bandwidth includes both the signal and idler modes, these two modes act effectively as a single mode and the DPA is a phase-sensitive amplifier.
On the contrary, if the idler frequency ($-\omega$) falls outside the measured frequency band, the idler mode acts as a noise mode and the DPA is a phase-preserving amplifier with phase-preserving photon number gain~\cite{caves:1982a,Eichler:2013kx}
\begin{equation}
	G_\omega = \abs{g_{S,\omega}}^2.
	\label{eq:phasePreservingGain}
\end{equation}
Hence, depending on the experimental details, the same system can act either as a phase-sensitive or phase-preserving amplifier. The same will hold true for the JPA, and we will thus consider both cases when characterizing the JPA properties as an amplifier in Sec.~\ref{sec::gainEta}.

Finally, in both operating regimes, the gain diverges ($D[\omega]=0$) at the parametric threshold~\cite{Laflamme:2011vn,Wustmann:2013uq}
\begin{equation}
	\lcrit  = \sqrt{\Delta^2 + \kappaTot^{2}/4}
	\label{eq:paramThresh}
\end{equation}
and large gain is obtained near but below this value.
Indeed, above the threshold spontaneous parametric oscillation effects will occur leading to the  generation of photons activated by vacuum and thermal fluctuations~\cite{Wilson:2010vn,Wustmann:2013uq}. In this work, we focus on parameter regimes where the DPA act as an amplifier and thus $\lambda < \lcrit$ is always considered.

\section{Higher-order corrections to the JPA: Comparison of pumping schemes}
\label{sec::LJPApumps}
In this section, we introduce the standard lumped-element JPA circuit
and consider three pumping schemes: the monochromatic current pump, the bichromatic current pump  and the monochromatic flux pump. We show how these amplifiers can be mapped back to the DPA model, and compare their respective advantages. Importantly, for each pumping scheme, we derive the leading nonidealities which cause deviations from the DPA model. The study of these deviations in the following sections constitutes the core of our results.

\subsection{JPA circuit}\label{subsec::LJPAcircuit}
As illustrated in Fig.~\ref{fig:circuit}(b), the lumped-element JPA is simply a capacitively shunted Josephson junction coupled to a transmission line~\cite{Yurke:1988kb,Hatridge:2011qf}.
The Hamiltonian of this standard circuit is
\begin{equation}
	\hat{H}_{\mathrm{JPA}} =  \frac{\hat{Q}^2}{2C} - E_J \cos \parO{ \frac{\hat{\phi}}{\varphi_0}},
\end{equation}
with $C$ the capacitance, $E_J$ the Josephson energy, $\varphi_0=\hbar/2e$ the reduced flux quantum, $\hat{Q}$ the charge and $\hat\phi = \int_{-\infty}^{t}\drm \tau \hat V(\tau)$ the generalized flux. As usual, the charge and the flux are conjugate quantum operators obeying
$[ \hat\phi \mathrm{,} \hat Q ] = i $. 

Expanding the cosine and introducing bosonic annihilation  
(creation) operator $\ah^{(\dag)}$, one obtains~\cite{Bourassa:2012fk}
\begin{equation}
	\hat{H}_{\mathrm{JPA}} = \omega_0 \ah^\dag \ah- E_J \sum_{n>1}^{\infty} \frac{\parO{-\ffpz^2}^n}{(2n)!} \parO{\ah+\ah^\dag}^{2n},
	\label{eq:HJPA_sum}
\end{equation}
with $\omega_0 = \sqrt{8 E_J E_C}$ the bare frequency of the resonator, $E_C = e^2/2C$ the charging energy and $\ffpz = 2 \sqrt{E_C/ \omega_0}$ the unitless zero point flux fluctuations.

As JPAs are usually weakly nonlinear devices, the next step is to perform a quartic potential approximation
 by keeping only the first correction to the standard harmonic oscillator
\begin{equation}
	\hat{H}_{\mathrm{JPA}} \approx 
	 \omega_0 \ah^\dag \ah
	 +   \frac{ \Lambda}{6} 	\parO{\ah+\ah^\dag }^{4}, \label{eq:hKerr_beforeRWA}
\end{equation}
with the Kerr coefficient $\Lambda = - E_J \ffpz^4 /4 = -E_C/2$. 
As the leading neglected correction is of order $ \Lambda^2/\omega_0$, this approximation is valid in the regime $\abs{\Lambda}/\omega_0 \ll 1$ which is relevant for typical JPA frequencies and Kerr nonlinearities corresponding to $\abs{\Lambda}/\omega_0 \sim 10^{-2}$ to $10^{-6}$~\cite{Bourassa:2012fk}.
We note that the Transmon qubit has the same circuit and Hamiltonian as a JPA but operates at larger Kerr nonlinearities~\cite{Koch:2007pb}.

In the rotating-wave approximation (RWA), also valid for $\abs{\Lambda} \ll \omega_0 $, we obtain the standard Kerr Hamiltonian 
\begin{equation}
	\hat{H}_{\mathrm{Kerr}} = 
	\tilde{\omega}_0 \ah^\dag \ah
	 + \Lambda \ah^{\dag 2} \ah^2, \label{eq:hKerr_afterRWA}
\end{equation}
with $\tilde{\omega}_0 = \omega_0 - 2\Lambda$ the renormalized oscillator frequency. By normal ordering and performing the RWA prior to the quartic potential approximation, corrections from all orders  renormalizes the resonator frequency and the Kerr nonlinearity~\cite{Leib:2012kx}. To keep the notation light, we neglect this simple renormalization of parameters in the present work.

While the Kerr Hamiltonian was obtained from the lumped-element circuit, the same Hamiltonian with renormalized parameters would apply for a distributed nonlinear resonator in the single-mode approximation~\cite{Bourassa:2012fk,Eichler:2013fk}, or a lumped-element JPA with additional linear inductances~\cite{Zhou:2014fk}. However, it is worth noting that, in both cases, the additional inductances can reduce the validity of the quartic potential approximation. See Ref.~\cite{Eichler:2013fk} for details.

As discussed in Sec.~\ref{sec::DPA}, in order for this system to act as a parametric amplifier a pump must modulate one of the parameters at twice the resonance frequency. We now consider three pumping schemes leading to such a modulation.

\subsection{Monochromatic current pump} \label{subsec::singlePump}

We first consider the standard current-pumped JPA~\cite{Yurke:1988kb,Yurke:1989fk,Yurke:1989kx,Castellanos-Beltran:2007ys,Hatridge:2011qf,Yurke:2006fk,Eichler:2013fk}. In this scheme, a single current pump near resonance with the oscillator is used. Due to the Josephson relations, the junction acts as a current-dependent nonlinear inductance with~\cite{Manucharyan:2007fk}
\begin{equation}
	\label{eq::nonlinearInductance}
	L_{\mathrm{NL}}(I) = L_J
	\parS{
		1 + 
		\frac{1}{6}
		\parO{
			\frac{I}{I_c}
		}^2 
		+ \cdots
	} .
\end{equation}
For a monochromatic current pump $I_p \propto \cos \wpump t$, the first nonlinear contribution to the inductance is the cosine squared, which leads to a modulation of the inductance at twice the pump frequency. 

Using standard circuit quantization techniques~\cite{Devoret:1995fk} and using the quantum optics language, the current pump is equivalent to adding a single-photon drive, such that the total Hamiltonian of the pumped circuit is
\begin{equation}
	\hat{H}_{1} = \hat{H}_{\mathrm{Kerr} } 
	+ \epsilon\, \erm^{-i \wpump t} \ah^\dag 	+  \epsilon^* \erm^{i \wpump t} \ah,
	\label{eq:h1pump}
\end{equation}
where $\epsilon$ and $\wpump$ are the pump amplitude and frequency.
In a frame rotating at the pump frequency, it is useful to eliminate the pump Hamiltonian using a displacement transformation, which leads to $\ah \rightarrow \alpha + \dhat$, with $\alpha$ the classical field and $\dhat$ the quantum fluctuations~\cite{Yurke:2006fk,Laflamme:2011vn}, see Appendix~\ref{app::doubleDisplacement} for details. In this displaced frame, the Hamiltonian takes the form
\begin{equation}
	\hat{H}'_{1}  = 
		\Dsp  \dhat^\dagger \dhat   + \frac{\lsp}{2} \dhat^{\dagger 2} + \frac{\lsp^*}{2} \dhat^2 
		+ \corrSinglePump,
	\label{eq::hSinglePump}
\end{equation}
with  the shifted detuning due to the cavity population
\begin{equation}
	\Dsp = \tilde{\omega}_0  + 4 \abs{\alpha}^{2} \Lambda  -\omega_p, 
\end{equation}
the effective parametric pump strength $\lsp = 2 \alpha^2 \Lambda$, and the classical field $\alpha$ obeying the nonlinear differential equation
\begin{equation}
	i\dot \alpha = \epsilon + \parO{\tilde{\omega}_0-\wpump +2 \Lambda \abs{\alpha}^2 -i \frac{\kappaTot}{2} }\alpha.
	\label{eq::alphaDot}
\end{equation}
In general, the steady state of this cubic equation can exhibit bifurcation physics~\cite{Yurke:2006fk,Manucharyan:2007fk}. However, for the current-pumped JPA, the bifurcation threshold coincides with the parametric threshold and in the context of parametric amplification, parameters are chosen to be below the bifurcation point~\cite{Laflamme:2011vn}. Thus,  in what follows, $\alpha(t)$ is always a single-valued function.

For simplicity, $\hat{H}'_{1}$ in Eq.~\eqref{eq::hSinglePump} was obtained by performing the quartic approximation before the displacement transformation. However, one can perform the displacement transformation on the full cosine potential, before performing the RWA and the quartic potential approximation. 
These corrections, which mainly shift the operating frequency and bifurcation point of the amplifier have been studied in details in Ref.~\cite{Kochetov:2015ab}.

The displaced Hamiltonian of Eq.~\eqref{eq::hSinglePump} also includes
the nonlinear corrections to the single pump scheme
\begin{equation}
	\corrSinglePump =  \mu  \dhat^{\dagger 2} \dhat + \mu^*\dhat^\dagger \dhat^2 
	+ \Lambda \dhat^{\dagger 2} \dhat^2,
\end{equation}
 with the cubic term coefficient $\mu = 2 \alpha \Lambda$ and the standard quartic Kerr coefficient $\Lambda$. 
These corrections originate from the displacement of the Kerr nonlinearity.
In order to obtain linearized equations, they can be neglected in the small quantum fluctuations limit ~\cite{Yurke:2006fk}.
 In the following sections, we will explore the validity of this approximation and show that it is valid in the low gain and low Kerr nonlinearity regime.

When neglecting the corrections $\corrSinglePump$, the displaced Hamiltonian $\hat{H}'_{\mathrm{1}}$ can be related to $\hat{H}_{\mathrm{DPA}}$ with the mapping
$\Dsp \rightarrow \Delta$ , 
$\lsp \rightarrow \lambda$, 
$\hat {d} \rightarrow \hat {a} $,
 and 
 $\dout = \aout - \sqrt{\kappa} \alpha  \rightarrow \aout$. 
Thus, in this linear regime, the monochromatic current-pumped JPA is equivalent to the DPA, but in a displaced frame. From the lab frame, this implies that while the output of a DPA is squeezed vacuum, the output of this current-pumped JPA is a displaced squeezed state.

\subsection{Bichromatic current pump}\label{subsec::doublePump}

\begin{figure}[tb]
	\includegraphics[width=\linewidth]{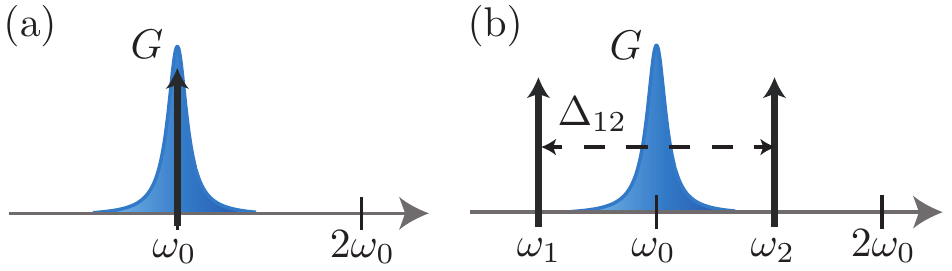}
	\caption{
	Relative position of the current pumps frequencies (black vertical arrows), and the frequency band of amplification (dark blue region) is shown for
	(a) monochromatic current pump with a single pump of frequency $\omega_{\mathrm{p}} \approx \omega_0$, and 
	(b) bichromatic current pumps with two pumps of frequencies $\omega_1 + \omega_2 \approx 2 \omega_0$.
	}
	\label{fig:frequencySpacePump}
\end{figure}

We now consider an alternative scheme with two current pumps of frequencies 
$\omega_1$ and $\omega_2$ that are chosen such that
$\omega_1+\omega_2 \approx 2\tilde{\omega}_0$~\cite{Kamal:2009uq}. As shown in Fig.~\ref{fig:frequencySpacePump}, while in the monochromatic case the pump is at a frequency close to the amplified signal [panel (a)], in the bichromatic case the pumps are separated spectrally from the signal [panel (b)]. 

Similar to Eq.~\eqref{eq:h1pump}, the starting Hamiltonian is 
\begin{equation}
		\hat{H}_{2} = 
	\hat{H}_{\mathrm{Kerr} }
	 + \sum_{n=1}^2 
	 \parS{
	 	\epsilon_n\, \erm^{-i \omega_{n} t} \ah^\dag 	
		 + h.c. 
	 },
	 \label{eq::HdoublePumpInitial}
\end{equation}
with $\omega_{1,2}$ and $\epsilon_{1,2}$ the frequencies and amplitudes of the two pumps. 
In order to remove the pumps in a similar way as in the monochromatic case, we consider two displacement transformations instead of one. This allows to  consider two classical fields, each rotating at one of the pump frequencies, and to separate them from the quantum fluctuations of the cavity mode possibly rotating at a third frequency.

Choosing a frame rotating at the average pump frequency $\Omega_{12} = (\omega_1 + \omega_2)/2 \approx \tilde{\omega}_0$, the double displacement transformation leads to 
\begin{equation}
	\ah \rightarrow \dhat\, \erm^{-i \Omega_{12} t} + \alpha_1\erm^{-i \omega_1 t} + \alpha_2 \erm^{-i \omega_2 t}.
\end{equation}
A more detailed and formal treatment of the transformation is presented in Appendix~\ref{app::doubleDisplacement}.
By applying this transformation, one obtains
\begin{align}
\begin{split}
\label{eq::bichromaticPumpHamiltonianAfterDisplacement}
	\hat{H}'_{2} &= 
	 \Dtdp \dhat^\dagger \dhat +
	  \frac{\ldp}{2} \dhat^{\dagger 2}+ \frac{\ldp^*}{2} \dhat^{ 2}
	+
	\hat H_{\mathrm{R}}
	+ \corrDoublePump
	,
\end{split}
\end{align}
with the shifted detuning 
\begin{equation}
	\Dtdp = \tilde{\omega}_0  + 4\Lambda \parO{\abs{\alpha_1}^2+\abs{\alpha_2 }^2}  - \Omega_{12},
\end{equation}
 and the effective parametric pump strength $\ldp = 4 \Lambda \alpha_1 \alpha_2$. 
 All the rotating terms are grouped in the Hamiltonian 
\begin{align}
	\hat H_{\mathrm{R}}
		&=
		8 \Lambda  \re{\alpha_1 \alpha_2^* \erm^{-i \Delta_{12} t}} 
		\dhat^\dagger \dhat
		\nonumber \\ &  +
		\parS{
			\Lambda \parO{\alpha_1^2 \erm^{-i \Delta_{12} t}+\alpha_2^2 \erm^{i \Delta_{12} t}} 
			\dhat^{\dagger 2}
			+ h.c.
		}
		\\ &  +
		\parS{
			 2 \Lambda \parO{\alpha_1 \erm^{-i \Delta_{12}t/2} + \alpha_2 \erm^{i \Delta_{12} t/2} }\dhat^{\dagger 2}\dhat
			+ h.c.
		},
		\nonumber
\end{align}
where $\Delta_{12} = \omega_1-\omega_2$ is the detuning between the two pump frequencies.
Finally, we define
\begin{equation}
	 \corrDoublePump = \Lambda \dhat^{\dagger 2}\dhat^2,
\end{equation}
 the nonlinear correction to the Hamiltonian.

The rotating Hamiltonian $\hat{H}_{\mathrm{R}}$ can be dropped by a RWA. 
Assuming symmetric pumps ($\alpha_1 \sim \alpha_2$), this RWA is valid for 
$\Delta_{12} \gg 8 \Lambda \alpha_1 \alpha_2 = 2\ldp$. 
Since the effective parametric pump strength is bounded by the parametric threshold $\lcrit$ defined in Eq.~\eqref{eq:paramThresh},
one can choose the detuning $\Delta_{12}$ in order to enforce the validity of the RWA.
For zero detuning in Eq.~\eqref{eq::bichromaticPumpHamiltonianAfterDisplacement} ($\Delta_2 = 0$), the RWA condition is simply $\Delta_{12} \gg \kappaTot$.
An advantage of the bichromatic pump compared to the monochromatic pump is thus that the cubic terms ($\mu \dhat^{\dagger 2}\dhat  + h.c.$) are now rotating and can be safely neglected.
This reduces the nonidealities of the amplifier, and implies that, with respect to the monochromatic current pump, the bichromatic current-pumped JPA acts as an ideal phase-sensitive amplifier for a larger parameter range (see Sec.~\ref{sec::intracavitySignatures}).

Again, in the small quantum fluctuations limit, we can neglect the higher-order correction $\corrDoublePump$. Under this approximation, the system is related to the DPA with the mapping $ \Dtdp \rightarrow \Delta$, $\ldp \rightarrow \lambda $, $\dhat \rightarrow \ah$ and
\begin{equation}
	\dout = \aout - \sqrt{\kappa} (\alpha_1 \erm^{-i \Delta_{12}t/2}+\alpha_2 \erm^{i \Delta_{12}t/2}) \rightarrow \aout.
\end{equation}
Whereas in the monochromatic case the pump leads to a displacement of the output field at the center frequency of the amplified band, 
in this case the two pumps lead to displacements far detuned from the band of interest. 
This spectral separation allows to filter the pumps without the need for a more involved pump cancellation scheme~\cite{Eichler:2013kx}.

\begin{table*}[tb]
	\caption{Comparison of pumping schemes properties, parameters, and leading Hamiltonian corrections.
	}
	\label{tab:comparingPumpingSchemes}
	\centering
	\begin{center}
	\begin{ruledtabular}
	\begin{tabular}{lccc}
		Pumping scheme
		 & Monochromatic  current
		 & Bichromatic  current  
		 & Monochromatic  flux  \\ \hline
	 Spectral separation & No & Yes & Yes  \\ 
	 Spatial separation & No & No & Yes  \\ 
	 Output state 
	 \footnote{
	 	We use the standard notation of quantum optics where $\hat D(\beta)$ and $\hat S(\xi)$ are respectively the displacement and squeezing operators~\cite{Barnett:1997fk}.
	 }
	 & 
	 $\hat D(\alpha)\hat S(\xi)\ket{0}$
	 & 
	 $\hat D(
	 \alpha_1 \erm^{i \Delta_{12}t}
	 +
	 \alpha_2 \erm^{-i \Delta_{12}t}
	 )
	 \hat S(\xi)\ket{0}$
	 &
	 $\hat S(\xi)\ket{0}$
	  \\ \hline
	 Effective parametric pump ($\lambda$)
	 	 & $2 \Lambda \alpha^2 $
	 	 & $4 \Lambda \alpha_1 \alpha_2$
	 	 & $4 \Lambda_{\mathrm{f}} \parO{E_J^{(1)}/ \omega_0}$
	 	  \\ 
	 Pump induced frequency shift ($\Delta^{\mathrm{(p)}}_{1,2,\mathrm{f}}$)
	 &
	 $4 \Lambda \abs{\alpha}^2 $
	 &
	 $4  \Lambda \parO{\abs{\alpha_1}^2+ \abs{\alpha_2}^2}$
	 &
	 $\omega_0 \parO{\sqrt{J_0(\df)}-1}$
	 \\ 
	 Relative shift ($\Delta^{\mathrm{(p)}}_{1,2,\mathrm{f}}/\omega_0$)
	 \footnote{
	 We introduce the compact notation
	 	$x = \abs{\alpha_1/\alpha_2}$, and $\tilde{Q} = Q \tan F $.
	 }
	 &
	 $ -\abs{\lsp/ \lcrit}/Q $
	 &
	 $ -\abs{\ldp/ \lcrit}\parO{x + x^{-1}}/2Q $
	 &
	 $-\abs{\lfp/ \lcrit}^2/2\tilde{Q}^2$
	\\
	 Corrections ($H_{(1,2,\mathrm{f})\mathrm{c}}$)& 
	 $\mu \dhat^{\dag 2}\dhat +\mu^* \dhat^{\dag }\dhat^2 $$+ \Lambda \dhat^{\dag 2}\dhat^2$  
	 & 
	$\Lambda \dhat^{\dag 2}\dhat^2$ & $\Lambda_{\mathrm{f}} \ah^{\dag 2}\ah^2$ 
	\end{tabular}
	\end{ruledtabular}
	\end{center}
\end{table*}
\subsection{Monochromatic flux pump}\label{subsec::fluxPump}
Current pumps are an indirect way to modulate the inductance of the JPA by using the nonlinearity of the Josephson junction. 
A well-studied alternative is to use an adjustable inductance that can be directly modulated. 
In superconducting circuits, this can be done by replacing a single Josephson junction by a SQUID, a flux-dependent nonlinear inductance. 
With this slight modification of the circuit, parametric amplification can be obtained by flux-pumping. This pumping scheme has been extensively studied both experimentally~\cite{Yamamoto:2008dp,Zhou:2014fk,Krantz:2013vn} and theoretically~\cite{Wustmann:2013uq}.
Here, we derive the Hamiltonian of the flux-pumped JPA similarly to Ref.~\cite{Wustmann:2013uq}, with an additional focus on higher-order corrections to the Hamiltonian.

Replacing the Josephson junction by a SQUID modifies the Josephson energy in the equations of Sec.~\ref{subsec::LJPAcircuit} such that~\cite{Tinkham:2004uq}
\begin{equation}
	E_J \rightarrow E_J \cos \parO{\frac{\Phi_x}{2\varphi_0}},
\end{equation}
where $\Phi_x$ is the external flux applied in the SQUID loop. 
In order to obtain parametric amplification, the external flux is modulated at frequency $\wpump \approx 2 \tilde{\omega}_0 $, with an additional static component chosen such that~\cite{Wustmann:2013uq}
\begin{align}
	\frac{\Phi_x}{2 \varphi_0} = F + \df \cos \omega_p t \,,
\end{align}
with $F$ the unitless static flux and $\df$ the modulation amplitude. 

In order to separate the harmonics of the pump, we Fourier expand the flux-dependent Josephson energy 
\begin{equation}
	E_J \cos \parO{F + \df \cos \wpump t}  = \sum_n E_J^{(n)} \cos(n \wpump t),
	\label{eq::fourierExpandFluxPump}
\end{equation}
where the coefficients of the expansion $E_J^{(n)}$ are given in Appendix~\ref{app::fluxPumpCoeff}.
In the relevant limit of small pump amplitude ($\df \ll 1 $), one obtains that the leading contribution of each Fourier coefficient is $E_J^{(n)} \propto \parO{\df/2}^n/n!$ and the expansion of Eq.~\eqref{eq::fourierExpandFluxPump} can be safely truncated.

Due to the flux modulation, the result of the RWA on Eq.~\eqref{eq:hKerr_beforeRWA} is modified and the Hamiltonian is
\begin{equation}
	\hat H_{\mathrm{f}} = \Dfp \ah^\dagger \ah + \frac{\lfp}{2} \parO{\ah^{\dagger 2} + \ah^2} 
	+ \corrFluxPump,
\end{equation}
 with the detuning $\Dfp = \tilde{\omega}_0- \omega_{p}/2$ and the effective parametric pump strength $\lfp = E_J^{(1)}\ffpz^2/2$. In the quartic potential approximation, the non-quadratic corrections to the Hamiltonian are now
 \begin{equation}
 \begin{split}
 	\corrFluxPump &= 
 	\Lfp \ah^{\dagger 2 }\ah^{2} 
 	-
 	\frac{E_J^{(1)} \ffpz^4}{12}\parO{  \ah^\dagger \ah^3 + \ah^{\dag 3}\ah}
 	\\ &\quad
 	- \frac{E_J^{(2)}\ffpz^4}{48}\parO{  \ah^4 + \ah^{\dag 4}},
 	\label{eq::corrFluxPump}
\end{split}
 \end{equation}
 with
 $ \Lfp= - E_J^{(0)} \ffpz^4 /4 = \Lambda J_0(\df) \cos F $ the Kerr nonlinearity renormalized by the flux modulation.
 
While the Kerr nonlinearity is essential for parametric amplification in the current-pumped cases, the parametric pump strength of the flux-pumped JPA is independent of this quantity. It is merely an artifact of the use of Josephson junctions to build a flux-dependent inductance.  
More explicitly, in the limits of a resonant pump $\Dfp \sim0$ and a small pump amplitude $\df \ll 1$, 
 the ratio of the parametric pump strength to the parametric threshold is
\begin{equation}
	\abs{\lfp/\lcrit } 
	\approx  \df \, Q \tan F ,
	 \label{eq::fluxPumpParametricStrength}
\end{equation}
with $Q = \omega_0/\kappaTot$ the JPA quality factor~\cite{Zhou:2014fk}. This ratio is independent of the Josephson energy or of the Kerr nonlinearity.

For standard JPA quality factors  $Q \sim$ 10-100, and static flux bias such that $\tan F \gtrsim 1$, Eq.~\eqref{eq::fluxPumpParametricStrength} implies that parametric pump strengths close to the parametric threshold can be obtained even in the small flux-pump limit $\df \ll 1$. 
Thus, the leading correction to the JPA Hamiltonian is the first term of Eq.~\eqref{eq::corrFluxPump} which is independent of $\df$. The other corrections, respectively linear and quadratic in $\df$, can be dropped in this small flux modulation limit. This implies that the leading higher-order correction to the flux-pumped JPA is the same as in the bichromatic current pump case ($\corrFluxPump \approx \corrDoublePump$) and thus the flux-pumped JPA Hamiltonian reads
\begin{equation}
	\hat H_{\mathrm{f}} \approx \hat H_{\mathrm{DPA}} + \corrDoublePump.
	\label{eq:ApproxHfluxPump}
\end{equation}
Again, in the limit of small quantum fluctuations, higher-order corrections can be dropped and the JPA related to the DPA model with the mapping $\Dfp\rightarrow \Delta$ and $\lfp\rightarrow \lambda$. 
Since there is no displacement transformation, this pumping scheme is more closely related to the DPA than the current-pumped JPAs. 

\subsection{Summary and comparison of pumping schemes}
\label{subsec::pumpSummary}
Table~\ref{tab:comparingPumpingSchemes} compares and  summarizes the pumping schemes reviewed in this section. It is divided in two parts:  the first presents general properties of each pumping scheme, while the second summarizes expressions for parameters, and leading higher-order corrections to the ideal DPA Hamiltonian.

The first distinction to make between these schemes is the spatial and spectral separation of the pump and signal. 
In the flux pump case, two distinct ports are used for the signal and the pump, 
while in the current pump cases the same input port is used for both the pump and the signal. Hence, while the output of the flux-pumped JPA is squeezed vacuum, the output of the current-pumped JPAs is displaced due to reflected pump field(s). 
In the bichromatic case these fields are far detuned from the amplified signal and can be filtered out prior to the measurement. On the contrary, in the monochromatic case the reflected pump is at the frequency where the amplifier gain is maximal. 
This implies that great care must be taken to either cancel the pump or separate it from the signal~\cite{Eichler:2013kx}.

Moreover, all three pumping schemes lead to a negative pump-induced frequency shift of the resonator. While in the case of the current pumps, the shift follows from the population of the nonlinear cavity by the pump field(s), in the case of the flux pump, the shift is a geometric effect due to the cosinusoidal dependence of the SQUID's Josephson energy on the external magnetic flux~\cite{Wustmann:2013uq,Krantz:2013vn}. 
These pump-induced detunings obey the relation 
\begin{equation}
	\abs{\Dfp^{(\mathrm{p})}} 
		\ll 
	\abs{\Dsp^{(\mathrm{p})}}
		\leq 
	\abs{\Ddp^{(\mathrm{p})}}
	, 
\end{equation}
where the $(\mathrm{p})$ superscript is used to note that we are considering the pump-induced shift.
Note that for the current pumps this shift scales as $Q^{-1}$, while in the flux pump case it scales as $Q^{-2}$ (see Table~\ref{tab:comparingPumpingSchemes}).

Finally, we note that while the leading higher-order correction is the same for the bichromatic current pump and the flux pump, the monochromatic current pump Hamiltonian has additional cubic corrections. This implies that, without any change to the actual amplifier circuit (fixed parameters), using a bichromatic current pump  instead of a monochromatic current pump decreases the nonidealities of the JPA.

\section{Signature of higher-order corrections in the intracavity field}
\label{sec::intracavitySignatures}
In order to evaluate the impact of the higher-order corrections discussed in the previous section, we numerically compute first and second order moments of the steady-state intracavity field.
These quantities show that, while the JPA acts as an ideal DPA in the low nonlinearity regime, the higher-order corrections can play a significant role in the presence of larger nonlinearities.
To give a more intuitive representation of the effect of the nonidealities on the amplifier state, we also calculate Wigner functions of the intracavity field.

The numerical results of this section are obtained by finding the steady-state solution of the master equation
\begin{equation}
	\dot {\hat \rho}  = -i \comm{\hat H_{\mathrm{DPA}} + \hat H_\alpha}{\hat \rho} + \kappaTot\, \mathcal{D} [\dhat]\hat \rho,
	\label{eq:ME}
\end{equation}
where $\hat H_\alpha = \corrSinglePump$, $\corrDoublePump$ are the higher-order corrections considered, and $\mathcal{D} [\dhat] \hat\rho = \dhat \hat\rho \dhat^\dag - (\dhat^\dag \dhat \hat \rho + \hat\rho \dhat^\dag \dhat)/2$ is the standard dissipation superoperator~\cite{Gardiner:2004fk,Walls:2008fk}.

 \begin{figure}[tb]
	\includegraphics[width=\linewidth]{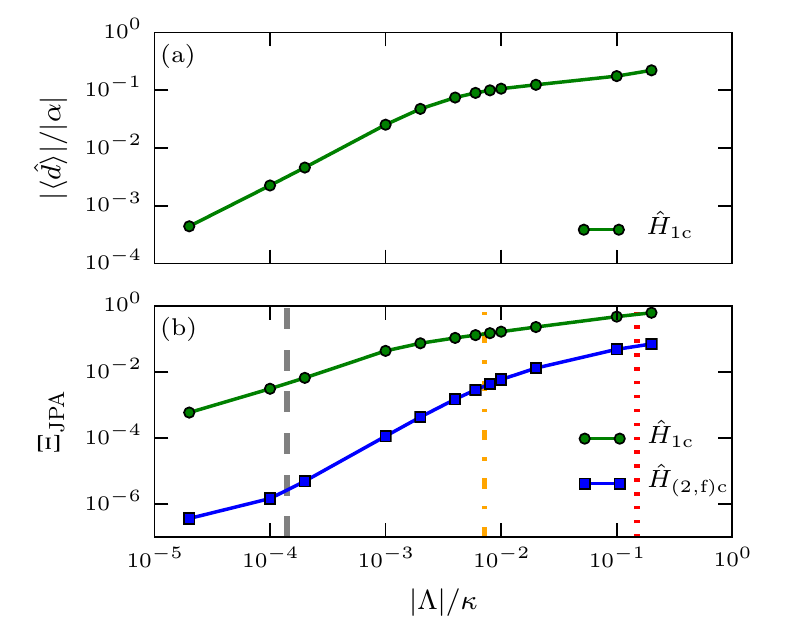}
	\caption{
		(a) Ratio of the displacement induced by higher-order corrections, to the steady-state solution of Eq.~\eqref{eq::alphaDot}. The Kerr correction  (bichromatic current pump and monochromatic flux pump) induces no displacement.
		(b) Deviation from standard DPA results as defined in Eq.~\eqref{eq:SigmaDPA}. From left to right, the vertical lines are approximate Kerr-nonlinearity for the experiments of Ref.~\cite{Eichler:2013kx}, Ref.~\cite{Hatridge:2011qf}, and Ref.~\cite{Murch:2013kx}. 
		Amplitude of the Kerr coefficient $\Lambda$ is varied for a fixed gain $G=16$~dB ($\lambda_{1,2} = 0.85\,\lcrit$) with $\Delta_{1,2}, \gamma=0$.
	}
	\label{fig:pumpDepletion}
\end{figure}
\begin{figure*}[tb]
	\includegraphics[width=0.8\linewidth]{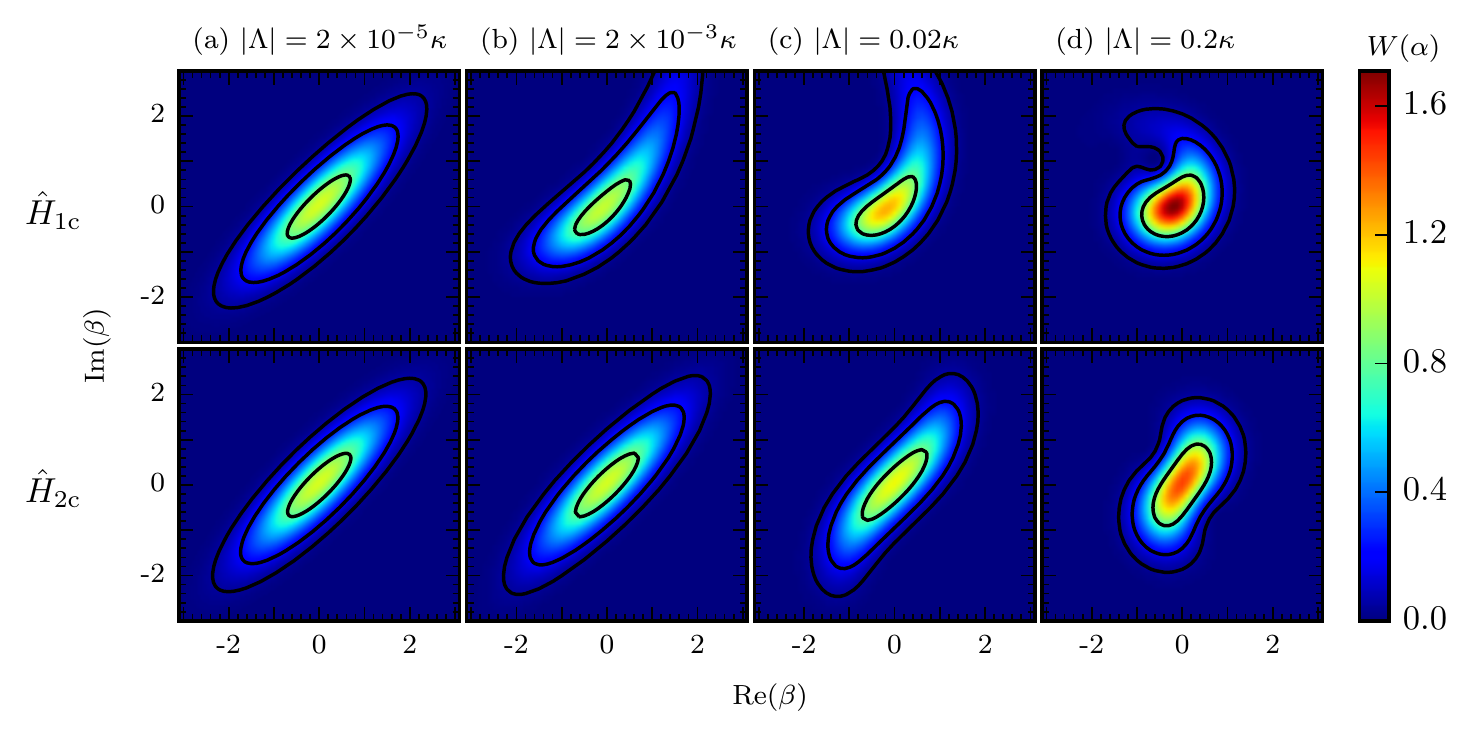}
	\caption{
	Steady-state Wigner function $W(\beta)$ of the JPA intracavity field for a gain 
	$G=16$~dB 
	($\lambda_{1,2,\mathrm{f}} = -0.85\,\lcrit$) with $\Delta_{1,2,\mathrm{f}}, \gamma=0$.
	Numerical result obtained via master equation simulation of the Hamiltonians of the monochromatic current pump Eq.~\eqref{eq::hSinglePump} in the displaced frame (top row) and of the bichromatic current pump Eq.~\eqref{eq::bichromaticPumpHamiltonianAfterDisplacement} after RWA which is equivalent to the monochromatic flux pump
	(bottom row).
	\label{fig:intracavityWigner}
	}
\end{figure*}

\subsection{Deviation from standard DPA results}
With the ideal DPA model, the first order moment of the intracavity field is $\langle \dhat \rangle =0$. However, in the case of the monochromatic current-pumped JPA, the cubic corrections act as an effective pump leading to an additional displacement of the field and thus a nonzero first moment.
To understand this effect, one can consider a mean-field treatment of the cubic corrections, where~\cite{Eichler:2013fk}
\begin{equation}
	\mu \dhat^{\dag 2 } \dhat + h.c.
	\approx
	  \parS{\parO{2 \mu   \overline{n} + \mu^* \overline{m} } \dhat^\dag    + h.c.}
\end{equation}
with $\overline{n} =\langle \dhat^{\dag} \dhat \rangle $, and $\overline{m} = \langle \dhat^2 \rangle$ the second-order moments of the JPA state. Under this approximation, the cubic terms act as an additional state-dependent pump and can be eliminated by a second displacement transformation.
Such a mean-field treatment was used to study pump depletion effects and the dynamic range of the JPA in Ref.~\cite{Eichler:2013fk}.

A mean-field solution would require the self-consistent solution for $\overline{n}$, $\overline{m}$.
Instead, we numerically find the steady-state of the master equation Eq.~\eqref{eq:ME}.
Fig.~\ref{fig:pumpDepletion}(a) shows the ratio of the displacement induced by the cubic terms, to the steady-state solution of Eq.~\eqref{eq::alphaDot} for the displacement field $\alpha$. While negligible in the low nonlinearity limit, the induced displacement becomes significant for larger Kerr nonlinearities. 
On the contrary and as expected from mean-field, no additional displacement is observed for the the bichromatic current and monochromatic flux pumps cases. 

We now consider signatures of the corrections in the second order centered moments $M_i = \langle \dhat^2 \rangle- \langle \dhat\rangle^2  $, and
 $N_i = \langle \dhat^\dag \dhat \rangle- |\langle \dhat \rangle|^2  $, where the subscript $i = $ DPA, JPA refers to the model that is used.
 It follows from Heisenberg uncertainty principle that, for any state, these moments obey the relation
 \begin{equation}
 	\abs{M_i} \leq \sqrt{N_i(N_i+1)},
 	\label{eq:MN}
 \end{equation}
 where the equality is only obtained for pure states~\cite{Gardiner:2004fk}.
From the analytical solution to the DPA model (see Sec.~\ref{sec::DPA} and App.~\ref{app::solvingDPA}), one obtains a value of $|M_{\mathrm{DPA}}|$ below this bound
 \begin{equation}
 	\abs{M_{\mathrm{DPA}}} = \sqrt{N_{\mathrm{DPA}} \parO{N_{\mathrm{DPA}}+1/2
 	}}.
 \end{equation}
 This result can be understood from the fact that, due to damping, the steady-state of the DPA is not a pure squeezed state~\cite{Collett:1984kl}.

 To quantify the deviation of the JPA moments from the expected results of an ideal DPA due to the corrections, we define the deviation
\begin{equation}
 	\Xi_{\mathrm{JPA}} 
 	= 1- \frac{\abs{M_{\mathrm{JPA}}}}{\sqrt{N_{\mathrm{JPA}}\parO{N_{\mathrm{JPA}}+1/2}}},
 	\label{eq:SigmaDPA}
 \end{equation}
where $M_{\mathrm{JPA}}$, $N_{\mathrm{JPA}}$ are the centered moments of the JPA intracavity field, here calculated numerically including higher-order corrections.
The deviation $\Xi_{\mathrm{JPA}}$ is zero in the case of an amplifier that maps exactly to a DPA (negligible higher-order corrections) and increases towards one as the corrections to the DPA Hamiltonian become important.

As observed in Fig.~\ref{fig:pumpDepletion}(b), the deviations increase with the amplitude of the Kerr nonlinearity. The additional cubic terms of $\corrSinglePump$ lead to larger deviation for the same parameters. Thus, by simply using two current pumps instead of one, the deviation from the expected results for a DPA are reduced by approximately two orders of magnitude. To put in context the range of Kerr nonlinearity considered, the vertical lines indicate approximate experimental parameters of three recent experiments with JPAs~\cite{Eichler:2013kx,Hatridge:2011qf,Murch:2013kx}. 
We note that, in practice, the smaller nonlinearity (dashed gray line) is obtained using junction arrays. Indeed, the Kerr nonlinearity with a junction array is inversely proportional to the square of the number of  junctions in the array~\cite{Castellanos-Beltran:2007ys,Eichler:2013fk}.

\subsection{Phase space representation}
In order to visually represent the deviation from the DPA, we present  in Fig.~\ref{fig:intracavityWigner} the Wigner function of the intracavity field for increasing Kerr nonlinearities (from left to right). 
For each value of $\abs{\Lambda}/\kappa$, we present results for the monochromatic current-pumped JPA (including cubic and quartic corrections $\corrSinglePump$) in the top panels and for the bichromatic current-pumped (quartic correction $\corrDoublePump$) or equivalently the flux-pumped JPA  in the bottom panels. In all cases, we consider a phase-preserving gain of $G= 16$~dB.
While in the low nonlinearity case presented in panels~(a) we observe a quadrature squeezed state for both type of corrections, in the increased nonlinearity of panels~(b) non-Gaussian signatures appear in the monochromatic pump case. Increasing even more the nonlinearity amplitude in panels~(c,d), both types of corrections lead to non-Gaussian signatures.

In the limit of large nonlinearity $\abs{\Lambda}/\kappa$, we observe that the shape of the Wigner function varies with the pumping scheme. 
In the monochromatic current pump case, we observe the so-called \textit{crescent} or \textit{banana}-shaped deformation of the distribution~\cite{dodonov2003theory}. Similar distributions have been observed in the transient dynamics of a driven Kerr nonlinear resonator.
In particular, the large nonlinearity regime $|\Lambda| \gg \kappa$ as been well studied both theoretically~\cite{Wilson-Gordon:1991qq}, and experimentally in superconducting circuits~\cite{Kirchmair:2013pi}.

In the case of a single Kerr-type correction, we observe a more symmetric ``S"-shaped Wigner function deformation.
In the case of the flux-pumped JPA, ``S"-shape features have been predicted theoretically and observed experimentally using a semi-classical analysis of the phase-dependence of the JPA response~\cite{Wustmann:2013uq,Bienfait:2016yq}.
This deformation of the Wigner function is also consistent with experimental studies imaging the JPA output field, see Sec.~\ref{sec::imaging}.

From the observed deformation of the Wigner function in panels (c,d), one can expect the higher-order corrections to limit the squeezing produced by a JPA. Such a saturation of squeezing has been observed experimentally~\cite{Murch:2013kx,Zhong:2013vn}, and will be discussed in more details in Sec.\ref{subsec:squeezingLevel}.
In addition, one can expect the output field of the JPA to exhibit significant non-Gaussian signatures for large gain and nonlinearity. Both the squeezing level and the non-Gaussian character of the output field are characterized more quantitatively in Sec.~\ref{sec::filteredOutput}.
More generally, the results of this section indicate that, for the same parameters, the additional cubic terms in the monochromatic current pump case lead to additional nonidealities limiting JPA performance. 

\section{Gain and quantum efficiency}
\label{sec::gainEta}

To characterize the effect of higher-order corrections on amplifier properties, in this section we numerically compute the low power gain and added-noise number of the JPA. From these, we obtain the phase-sensitive and phase-preserving quantum efficiency of the amplifier.

\subsection{Phase-sensitive and phase-preserving gain}\label{subsec:gain}
The gain is computed by considering the linear response of the JPA to a narrow-band signal probe or, in other words, the same approach as is used experimentally.
More concretely, starting from the steady-state solution of the JPA master equation Eq.~\eqref{eq:ME}, we add to the Hamiltonian the probe field 
\begin{equation}
	\hat H_{\mathrm{probe}} = \epsilon_{\mathrm{probe}} \ah^\dag + \epsilon_{\mathrm{probe}}^{*} \ah,
\end{equation}
with $\epsilon_{\mathrm{probe}}$ the drive amplitude, and find the new steady-state under this weak drive.
From the displacement of the cavity field generated by the probe, one can calculate the output field response.  
As is already made clear from Eq.~\eqref{eq::DPA_inputOutput}, the gain is a function of the probe frequency.
Using a time-independent probe, we compute the JPA gain at the center frequency $\omega=0$.

As discussed in Sec.~\ref{sec::intracavitySignatures}, the cubic corrections in the Hamiltonian of the monochromatic current-pumped JPA also induce a displacement of the cavity field.
This additional displacement modifies the bifurcation point of the system, something that can lead to instabilities in the numerical calculations. 
Hence, this section considers only the effect of a Kerr-type correction, which is the leading correction for the bichromatic current pump and the monochromatic flux pump. In the case of the monochromatic current pump, the additional cubic terms could lead to corrections which, following the results of the previous sections, would appear at lower gain and nonlinearity than those due to the quartic term.

To study the linear response regime, we limit our analysis to a low-power probe where $\epsilon_{\mathrm{probe}} \ll \kappa$.
While an analysis of the dependence of the gain on the probe amplitude would allow to calculate the dynamic range of the JPA~\cite{Eichler:2013fk,Kochetov:2015ab}, this is beyond the scope of this paper.

Using the input-output relation Eq.~\eqref{eq:boundary}, the displacement generated by the probe and the input field amplitude
	$\average{\ain} = {i \epsilon_{\mathrm{probe} }}/{\sqrt{\kappa}}$,
we can compute the gain matrix $\Grm$ whose elements are defined by the linear input-output equation
\begin{equation}
	\matO{\hat X_{\mathrm{out}} \\ \hat P_{\mathrm{out}}}
	=
	\matO{\Grm_{11} & \Grm_{12} \\ \Grm_{21}& \Grm_{22}}
	\matO{\hat X_{\mathrm{in}} \\ \hat P_{\mathrm{in}}},
	\label{eq:quadratureLinearInputOutput}
\end{equation}
where we have defined the standard quadratures
	$\hat X =  \parO{\ah+\ah^\dag}/\sqrt{2}$,
	and
	$\hat P =  i\parO{\ah^\dag-\ah}/\sqrt{2}$.
By considering separately the response to two probes with orthogonal phases, we can calculate all elements of the gain matrix.

To obtain the phase-preserving gain, we express the above quadrature input-output relation in terms of field operators. Using Eqs.~\eqref{eq::DPA_inputOutput} and \eqref{eq:phasePreservingGain}, we then find that the JPA phase-preserving gain is related to the phase-sensitive gain matrix by
\begin{equation}
 	\tilde{G}
 	= 
 	\frac{1}{4}\abs{\Grm_{11}+ \Grm_{22}+ i \parO{\Grm_{21}-\Grm_{12} }}^2.
 	\label{eq:gainPP}
 \end{equation} 

\begin{figure}[tb]
	\includegraphics[width=\linewidth]{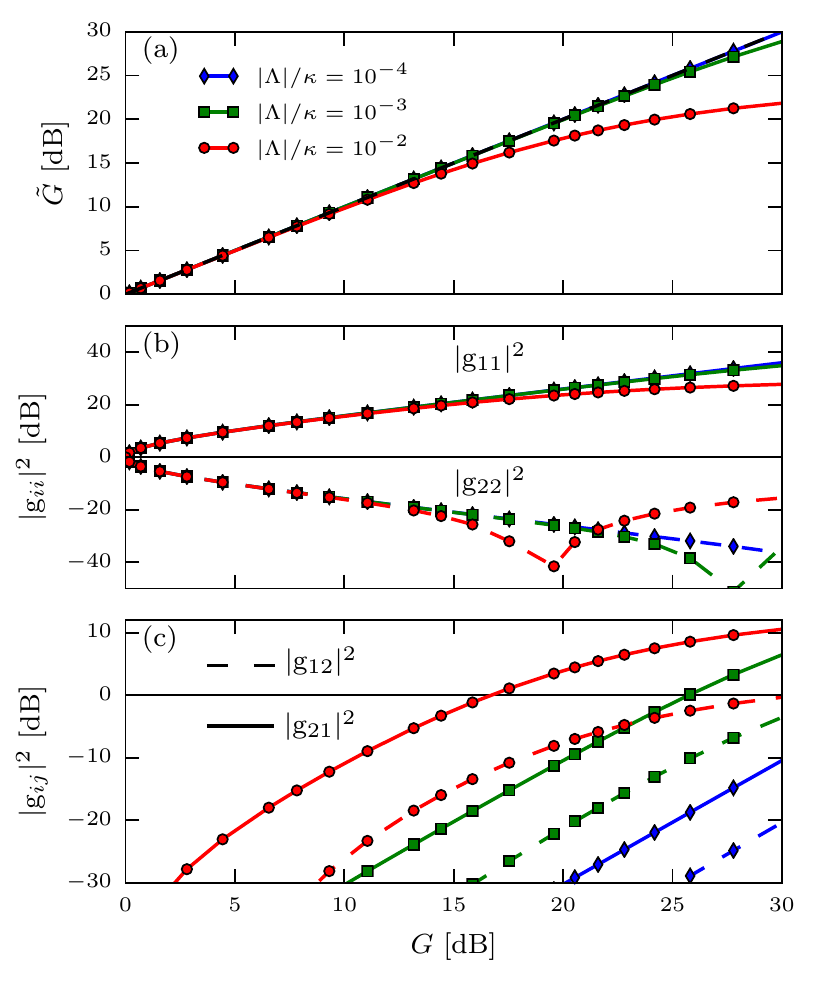}
	\caption{
	(a) Numerical phase-preserving gain for a JPA with Kerr-type correction $\hat H_{2c}$ calculated from Eq.~\eqref{eq:gainPP} versus gain  for the DPA calculated from Eq.~\eqref{eq:phasePreservingGain}. 
	(b) Norm of the diagonal elements $\Grm_{11}$ (solid curves) and $\Grm_{22}$ (dashed curves) of the phase-sensitive gain matrix defined at Eq.~\eqref{eq:quadratureLinearInputOutput}. 
	(c) Off-diagonal elements $|\Grm_{21}|^2$ (solid curves) and $|\Grm_{12}|^2$ (dashed curves). 
	For all curves, the gain is calculated for $\omega=0$, $\Delta=0$, and $\gamma=0$.
	}
	\label{fig:gain}
\end{figure}
Fig.~\ref{fig:gain} presents the phase-preserving gain, as well as the elements of the phase-sensitive gain matrix for a JPA with increasing Kerr nonlinearities. 
These quantities are plotted as a function of the phase-preserving gain calculated from Eq.~\eqref{eq:phasePreservingGain} for an ideal DPA. 
As expected, Fig~\ref{fig:gain}(a) shows that, in the low nonlinearity regime (blue diamonds), the JPA phase-preserving gain is equal to the ideal DPA gain and the corrections are negligible. As the Kerr nonlinearity increases (green squares and red circles) the JPA nonidealities result in a decreased gain, with deviations increasing with gain.

Fig~\ref{fig:gain}(b) presents the diagonal elements of the phase-sensitive gain matrix, while Fig~\ref{fig:gain}(c) shows the off-diagonal elements. In the low gain regime, as expected for a phase-sensitive amplifier, the diagonal elements are inversely proportional with one quadrature amplified and the other attenuated. 
In this regime, the gain matrix is diagonal. 
When the off-diagonal terms become significant, the attenuation coefficient $\Grm_{22}$ starts to increase, deviating significantly from the expected behavior of an ideal phase-sensitive amplifier.

For higher gain and nonlinearities, the gain matrix is not symmetric and cannot be diagonalized with orthogonal eigenvectors.
In this regime, the amplification process mixes quadratures. 
 Noting that a similar effect occurs for a DPA with pump-cavity detuning~\cite{Laflamme:2011vn}, we can interpret the effect of the nonlinearity as a gain-dependent detuning of the system. Hence, for a given gain choosing a slightly different detuning could mitigate higher order effects and reduce quadrature mixing. This implies that when higher-order corrections are important, the optimal phase and frequency of operation of the JPA is gain-dependent.
 The interplay of the optimal frequency of operation and nonlinear corrections has recently been the subject of experimental investigation in a similar device~\cite{hatrige2017}.

\subsection{Phase-preserving quantum efficiency}

\begin{figure}[tb]
	\includegraphics[scale=0.28]{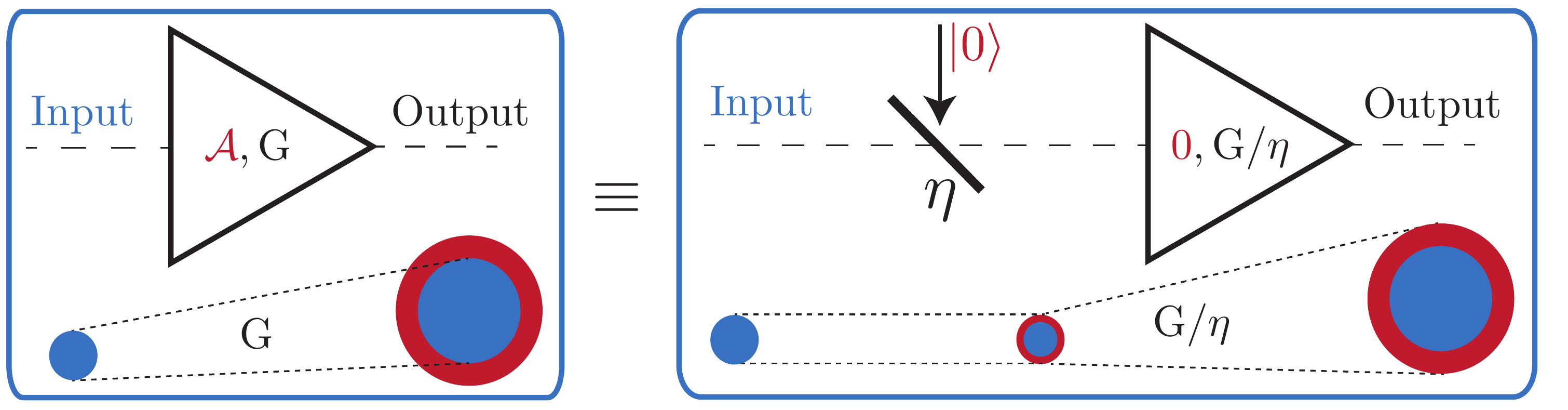}
	\caption{
		Schematics of the quantum efficiency definition as a beam splitter with transparency $\eta$ and input vacuum noise. The blue discs represent signal noise before and after amplification, while red discs represent noise added by the amplification process.
	}
	\label{fig:beamSplitterEta}
\end{figure}

To characterize the effect of Kerr-type correction on the JPA noise properties, we now evaluate its added-noise and quantum efficiency. As illustrated in Fig.~\ref{fig:beamSplitterEta}, the quantum efficiency $\eta$ can be interpreted as the transparency of a fictitious beam splitter added at the input of a noiseless amplifier in order to model the 
noise added by the amplification as additional input vacuum noise~\cite{Leonhardt1994,Mallet:2011fk}.

With this picture in mind, we define the quantum efficiency $\eta$ such that the input-output field fluctuations are related by 
	\begin{equation}
		\langle |\aout|^2 \rangle = \frac{ G }{\eta} \parS{  \parO{ 1- \eta } \frac{ 1 }{2} + \eta \langle|\ain|^2 \rangle   },
		\label{eq:inOutFluctuations}
	\end{equation}
with $\average{\abs{\cdot}}$ the symmetrized fluctuations of an operator
\begin{equation}
	\langle|\Omh|^2 \rangle = \frac{1}{2} \langle\Omh^\dag \Omh + \Omh \Omh^\dag\rangle.
	\label{eq:fluctuations}
\end{equation}
Note that the case $\eta=0$ corresponds to having no output signal and is therefore not relevant here. The first term of Eq.~\eqref{eq:inOutFluctuations} is the added noise due to the amplification process, represented as vacuum fluctuations in Fig.~\ref{fig:beamSplitterEta}, while the second term is fluctuations in the input signal.
It is useful to express Eq.~\eqref{eq:inOutFluctuations} in a simpler form
\begin{equation}
	 \langle |\aout|^2 \rangle  = G\parO{\Am+ \langle |\ain |^2  \rangle },
	\label{eq::phasePreservingNoise}
\end{equation}
where $\Am = (1-\eta)/2\eta$ is the added noise referred to the input. 
Using the inequality 
$ \langle|\Omh|^2 \rangle
	\geq \langle [ \Omh ,  \Omh^\dag ]\rangle|/2$,
one can derive the well-known quantum limit for a phase-preserving amplifier~\cite{caves:1982a}
\begin{equation}
	\Am \geq \frac{1}{2}\parO{1 - \frac{1}{G}},
	\label{eq:ql} 
\end{equation}
which simplifies in the large gain limit to $\Am \geq 1/2$.

Using these definitions, the quantum efficiency can be expressed as
\begin{equation}
	\eta = \frac{1}{1+ 2\Am} \leq \frac{G}{2G-1},
	\label{eq:eta}
\end{equation}
where we have used the quantum limit of Eq.~\eqref{eq:ql}.
This inequality implies, in the large gain limit, that the bound on the quantum efficiency of a phase-preserving measurement is $\eta \leq \frac{1}{2}$. This result simply reflects the well-known fact that ideal phase-preserving amplification can be obtained by the use of two ideal phase-sensitive amplifiers and a beam splitter which adds vacuum fluctuations~\cite{caves:1982a}.

Note that the alternative definition $\overline{\eta} =  1/(1+ \overline{N}) =2\eta$ of the quantum efficiency is also found in the literature, with $\overline{N} = \Am -1/2$ a number of added noise photons. 
This definition relates the amplifier performance to an ideal phase-preserving amplifier instead of a noiseless amplifier. However, this expression implicitly assumes the large gain limit of Eq.~\eqref{eq:ql} and therefore overestimates the quantum efficiency in the low gain regime where $1/G$ cannot be neglected.
 In this work, we will consider the definition of Eq.~\eqref{eq:eta}, as this allows to treat phase-preserving and phase-sensitive amplification on the same footing and is independent of gain.

\begin{figure}[tb]
	\includegraphics[width=\linewidth]{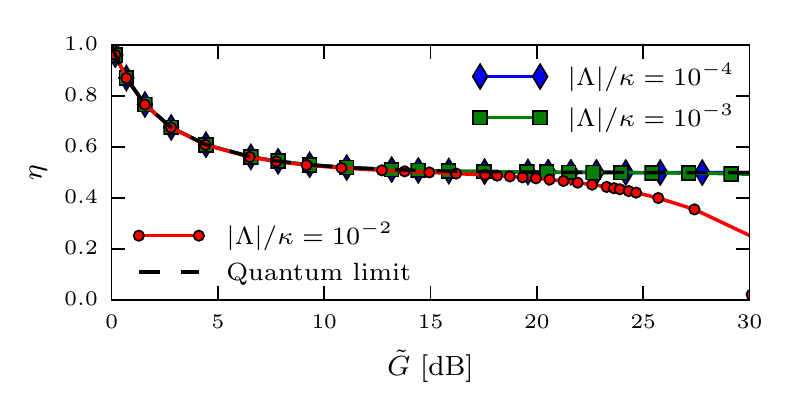}
	\caption{
	Phase preserving quantum efficiency ($\eta$) as a function of gain for a JPA with Kerr correction.
	The two lower Kerr nonlinearity cases considered $|\Lambda|/\kappa= 10^{-4}$ (blue diamonds) and $|\Lambda|/\kappa= 10^{-3}$ (green squares) are near quantum-limited and, as a result, are difficult to resolve from the quantum limit (dashed black curve) for the full range of parameters considered.
	For all curves, the quantum efficiency is calculated for $\omega=0$, $\Delta=0$, and $\gamma=0$.}
	\label{fig:phasePreservingEta}
\end{figure}
Fig.~\ref{fig:phasePreservingEta} shows the quantum efficiency in the phase-preserving case as a function of gain and for three Kerr nonlinearities. In the low gain regime, all curves are equal to the quantum limit. For higher gains, the Kerr nonlinearity leads to a decreased quantum efficiency. These results are obtained by numerically calculating the gain and the output field spectrum of the amplifier, with
\begin{equation}
	 \langle | \aout| ^2 \rangle  = n_{\mathrm{out}}[0] +\frac{1}{2}.
\end{equation}
In this calculation, we neglect any bandwidth or detuning effect and consider the zero frequency component of the spectrum. 

Surprisingly, even for a significant phase-preserving gain $\tilde{G}=20$~dB,  and Kerr nonlinearity $|\Lambda| = 0.01 \kappa$, the JPA with Kerr-type corrections is nearly quantum-limited without any tuning of parameters. On the contrary, we will show in the following section that the same Kerr-type correction strongly influences the quantum efficiency of a phase-dependent measurement. In that case, a careful choice of the phase of operation of the JPA is essential in order to obtain near quantum-limited amplification.

\subsection{Phase-sensitive quantum efficiency \label{subsec:phaseEta}} 
We now generalize the concepts of the previous section to the case of phase-sensitive amplification. Contrary to the simpler case of phase-preserving amplification, the quantum efficiency is not a single number, but rather a function of the measurement phase $\theta$. Hence, one must choose the measured quadrature in order to maximize quantum efficiency.

For the ideal DPA, no noise is added independently of the phase considered and $\eta(\theta)=1$ for all $\theta$. In that specific case, the quantum efficiency of the full measurement chain~\cite{pozar1997} is maximized for the measurement phase $\theta_m$ which maximizes the gain $\Grm_{11}$.
However, when including a Kerr correction, we have shown in Sec.~\ref{subsec:gain} that the gain matrix becomes non-symmetric. In that case, the nonzero $\mathrm{g}_{12}$ leads to quadrature mixing during the amplification, which can be seen as a source of added noise.
Thus, we will show that contrary to the ideal DPA, in the presence of a Kerr correction, the maximal quantum efficiency is not obtained by maximizing $\Grm_{11}$  but rather by minimizing $\Grm_{12}$. We note $\theta_0$ the phase of the quadrature which minimizes the mixing of noise with the signal. 

More formally, in order to characterize the field fluctuations, 
we define the matrix of the symmetrized moments as the matrix analog of Eq.~\eqref{eq:fluctuations}~\cite{caves:1982a}
\begin{equation}
	\sigma_j = \matO{
	\langle\Delta \hat X_{j}^2 \rangle
	& \frac{1}{2} \langle\{\Delta \hat X_{j} \,\mathrm{,}\, \Delta \hat P_{j} \}\rangle
	\\
	\frac{1}{2} \langle \{\Delta \hat X_{j} \,\mathrm{,}\, \Delta \hat P_{j} \}\rangle
	 & 
	 \langle\Delta \hat P_{j}^2 \rangle 
	},
\end{equation}
with $j= \mathrm{out,in}$, and $\{ \hat A \,\mathrm{,}\, \hat B \} = \hat A \hat B+\hat B \hat A$ the anticommutator.
Similarly, the added-noise matrix is defined through a generalization of Eq.~\eqref{eq::phasePreservingNoise} 
\begin{equation}
	\sigma_{\mathrm{out}} =  \Grm \parO{ \sigma_{\Am} +  \sigma_{\mathrm{in}}} \Grm^\mathrm{T}.
\end{equation}
We note that the product of the diagonal elements of this matrix are bounded by the quantum limit to amplification~\cite{caves:1982a}  
\begin{equation}
	\sigma_{\Am11} \sigma_{\Am22} \geq  \frac{1}{4}\abs{1 - \frac{1}{\Grm_{11}\Grm_{22}}}^2.
\end{equation}
For the standard lossless DPA, we obtain noiseless amplification since $\Grm_{22} = 1/\Grm_{11}$ and $\sigma_{\Am11} = \sigma_{\Am22} = 0$.

In the presence of Kerr correction, to make explicit the choice of the measurement phase, we define the phase-dependent added noise
\begin{equation}
	\Am(\theta) = \parS{R^{\mathrm{T}}(\theta) \sigma_\Am R(\theta)}_{11}
	\label{eq:ATheta}
\end{equation}
as the first diagonal element of the rotated added-noise matrix, with $R(\theta)$ the counter-clockwise
orthogonal rotation matrix.
From this, we define the phase-dependent quantum efficiency
	$\eta(\theta) = 1/[1+2 \Am(\theta)] \leq 1$.
While we focus in the following on the diagonal element of the added-noise matrix, with $\Am(\theta)$ characterizing the noise added to the amplified quadrature, in general $\sigma_\Am$ is non-diagonal and the amplification can lead to added cross-correlations between the quadratures.

\begin{figure}[tb]
	\includegraphics[width=\linewidth]{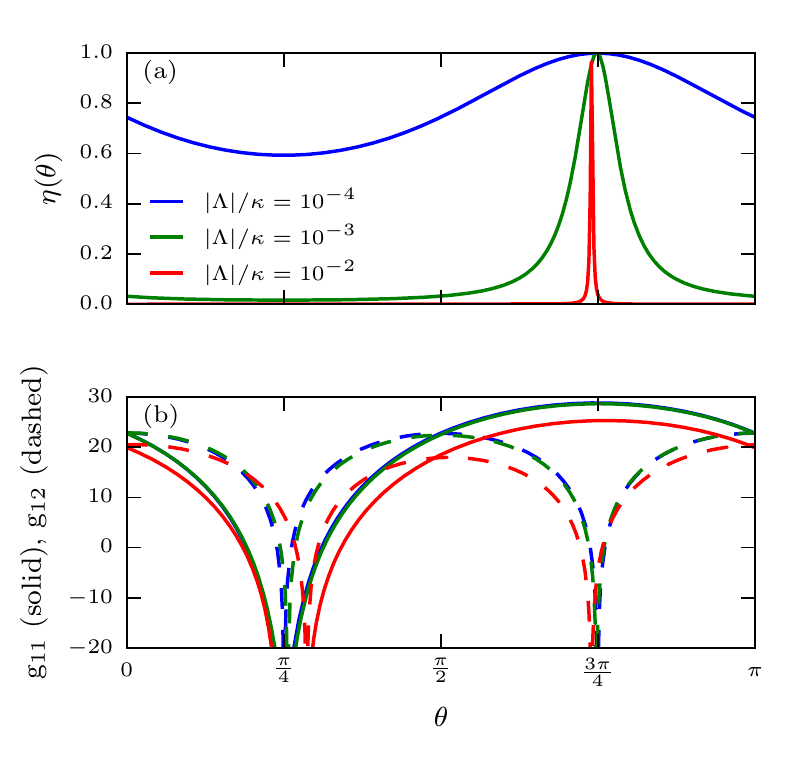}
	\caption{
	(a) Phase-sensitive quantum efficiency as a function of the phase of the measurement quadrature. 
	(b) Matrix elements of $\Grm$ ($\Grm_{11}$ solid curve, $\Grm_{12}$ dashed curve) as a function of the quadrature phase $\theta$. 
	All data corresponds to a photon number gain $G_{\mathrm{JPA}} = 23$~dB for a JPA with Kerr correction, and$\omega=\Delta =\gamma=0$.
	}
	\label{fig:etaTheta}
\end{figure}

Fig.~\ref{fig:etaTheta} shows the quantum efficiency and gain as a function of the quadrature phase $\theta$ for increasing values of $|\Lambda/\kappa|$. 
In Fig.~\ref{fig:etaTheta}(a), we observe that the quantum efficiency oscillates with $\theta$ and becomes increasingly peaked around a value of the phase with increasing nonlinearity. For a fixed nonlinearity, increasing the gain leads to the same effect (not shown).
We note that the position of the peak in the quantum efficiency correlates with a dip in the off-diagonal matrix element $\Grm_{12}$ shown in Fig~\ref{fig:etaTheta}(b). 
This dip is shifted to a narrower range of phases as the Kerr nonlinearity is increased. This is in agreement with Fig.~\ref{fig:gain}(c) which showed that, for a fixed quadrature phase, increasing the nonlinearity leads to larger $\Grm_{12}$ and thus requires a larger phase correction.
Thus, as expected the quantum efficiency is maximized when the noise added through quadrature mixing is minimized. 

\begin{figure}[tb]
	\includegraphics[width=\linewidth]{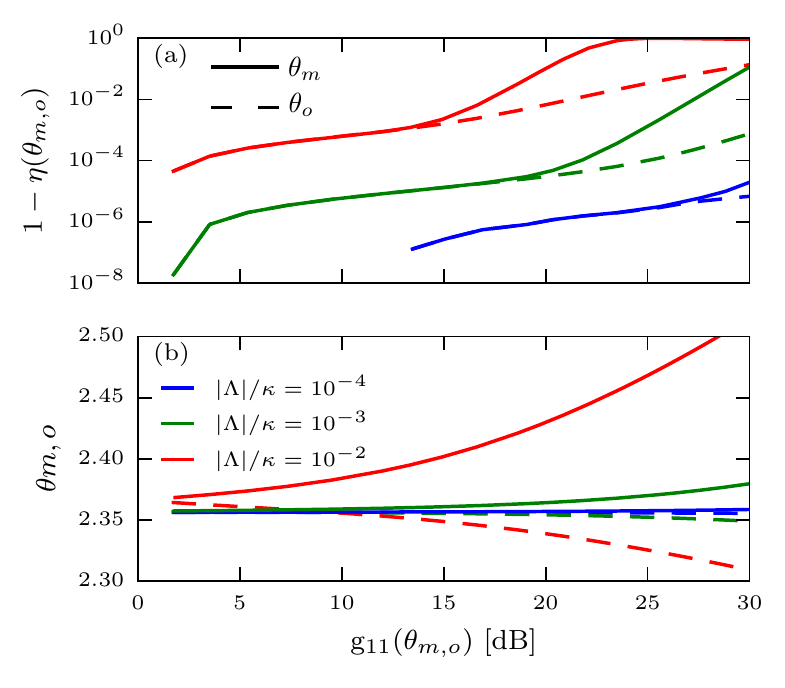}
	\caption{
		(a) Reduction in the phase-sensitive quantum efficiency due to Kerr-type corrections. Solid curves corresponds to the phase of maximal gain $\theta_m$, while dashed curves corresponds to the phase $\theta_o$, where the cross-gain $\Grm_{12}$ is minimal.
		(b) Shift in rad of the phases $\theta_m$ and $\theta_o$ as a function of gain. 
	}
	\label{fig:phaseSensitiveEta}
\end{figure}
Fig.~\ref{fig:phaseSensitiveEta} presents $1-\eta(\theta)$ as a function of phase-sensitive gain for increasing Kerr nonlinearities. To further illustrate the importance of choosing the appropriate phase, results for both the optimal phase $\theta_o$ of minimal cross-gain $\abs{\Grm_{12}} $ (dashed curves), and for the non-optimal phase $\theta_m$ of maximal gain ($|\Grm_{11}|$, solid curves)  are shown. 
As expected, the quantum efficiency is reduced [larger $1-\eta(\theta)$] for increased gain and nonlinearities. 
Strikingly, when considering a gain of 25~dB and $|\Lambda/\kappa|=10^{-2}$, the quantum efficiency at phase $\theta_o$ is around 0.9, while it is closed to zero for $\theta_m$. Fig~\ref{fig:phaseSensitiveEta}(b) shows that this dramatic difference in performance is obtained with less than $\sim0.2$~rad (12 degrees) difference in phase.

These results points to both the Kerr nonlinearity and the non-optimal choice of measurement phase as a possible explanation for a series of experiments that have reported smaller than expected quantum efficiencies for JPAs~\cite{Murch:2013uq,Murch:2013kx,Vijay:2012uq,weberThesis}. A more detailed experimental study of the quantum efficiency as a function of both detuning and measurement phase could confirm these results and would allow for a better understanding of these nonlinear effects.

\section{Output field characterization}
\label{sec::filteredOutput}
Following the results of the previous sections, one can expect to also find signatures of the nonidealities in the JPA output field, including in squeezing experiments. In this section, we compute moments of the output field which allow to characterize the JPA as a source of squeezed light. 
Following the results of Sec.~\ref{sec::intracavitySignatures} and more generally for Kerr cavities~\cite{dodonov2003theory,Kirchmair:2013pi}, we expect the higher-order corrections to lead to a non-Gaussian output field. 
To verify this, we calculate third and fourth order cumulants which reveal departure from gaussianity. 

Throughout this section, we compare our numerical results to an experimental characterization of the output field of a flux-driven JPA. Moments of the JPA output field are obtained using a single-path reconstruction method~\cite{Eichler_PRL_2011}, while a full output field imaging is obtained from a deconvolution technique. Details of the experimental setup and methods are presented in Appendix~\ref{App::exp}.

\subsection{Filtered output field : Definition and numerical technique}
In order to compare our numerical results to experimental data, we consider the finite bandwidth of the measurement chain in our calculations. 
To this end, we define the filtered output field $\hat D(t)$ as the convolution of the full output field $\aout$~\cite{Silva:2010mi}:
\begin{equation}
	\hat D(t) = f \star \aout (t)  
	= \int_{-\infty}^{\infty} \mathrm{d}\tau \, f(t-\tau) \aout(\tau),
\end{equation}
with a filter function $f$
normalized such that $\int_{-\infty}^\infty \! \drm t \, \abs{f(t)}^2  =1$, in order to ensure standard bosonic commutation relations $[\hat D \,, \, \hat D^\dag] =1$.
The moments of this field can be evaluated by numerical integration of correlation functions using the quantum regression formula~\cite{Gardiner:2004fk}. The details of the numerical technique are presented in Appendix~\ref{app::filteredMoments}. We note that the complexity of the calculation scales exponentially with the order of the moment considered~\cite{Silva:2010mi}. As a result, fourth order moments are at the limit of our current computational capacities. Fortunately, this is sufficient to characterize the departure from ideal Gaussian behavior.
Unless otherwise specified, the filter used is a time-domain boxcar filter of 256~ns (bandwidth of approximately 4~MHz), which coincides with the experimental method.

\subsection{Squeezing level}\label{subsec:squeezingLevel}
In order to characterize the squeezing produced by the JPA, we calculate the filtered output field squeezing level defined as~\cite{Zhong:2013vn}
\begin{equation}
	S_f = 
			\frac{
		 \langle \Delta \hat X^2_{\mathrm{Vac}} \rangle 
		}{ 
		\langle \Delta \hat X^2_{\mathrm{Min} }  \rangle
		}
	,
	\label{eq:sqLevel}
\end{equation}
with
$ \langle \Delta \hat X^2_{\mathrm{Vac}} \rangle  =1/2 $
the variance of the vacuum state, and
\begin{equation}
	 \langle \Delta \hat X^2_{\mathrm{Min} }  \rangle
	=N_f + \frac{1}{2} - \abs{M_f}
	,
\end{equation}
the minimal variance of the filtered field. Here, we have defined the centered moments of the filtered output field $N_f = \langle \hat D^\dag \hat D \rangle-|\langle \hat D \rangle |^2 $ and $M_f = \langle \hat D^2 \rangle-\langle \hat D \rangle^2$.

\begin{figure}[tb]
	\includegraphics[width=\linewidth]{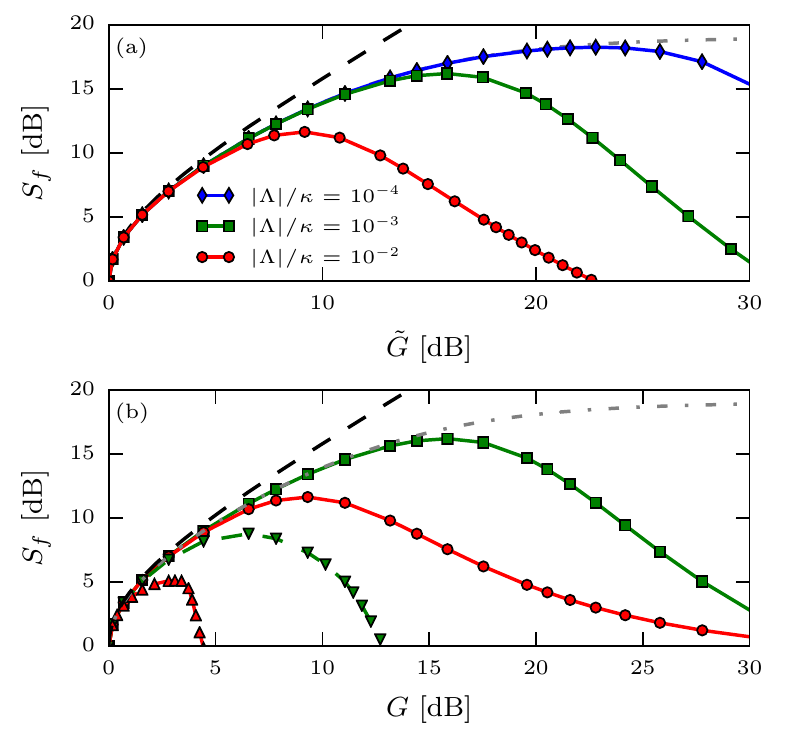}
	\caption{
	(a) Squeezing level of the filtered output field ($S_f$) versus numerical calculation of the gain using Eq.~\eqref{eq:gainPP}. 
	The filter is a time-domain boxcar filter of length $256$~ns.
	Dashed black curve (dotted-dashed gray curve) is the maximal squeezing level of a DPA without (with) the filter.
	Solid curves are numerical results including 
	$\corrDoublePump \approx \corrFluxPump$ corrections.
	(b)
	Comparison of the squeezing level with $\corrSinglePump$ corrections (dashed curves) or $\corrDoublePump \approx \corrFluxPump$ corrections  (solid curves)  as a function of the photon number gain calculated without corrections [\eqref{eq:phasePreservingGain}].
	(All panels : $\kappa/2\pi=50$~MHz, $\Delta=\gamma=0$.)
	}
	\label{fig:gainAndSqueezingNum}
\end{figure}

Fig.~\ref{fig:gainAndSqueezingNum}(a) shows the filtered squeezing level $S_f$ as a function of phase-preserving gain for a JPA with Kerr correction.
For an ideal JPA, the squeezing level of the center frequency (Dirac-delta filtering) increases with gain without bound (dashed black curve).
On the other hand, even for an ideal JPA (no higher-order corrections), the filtered squeezing level saturates for a finite-bandwidth filter  (dotted-dashed gray curve).
%
Indeed, as the gain increases, the squeezing bandwidth is reduced and eventually becomes smaller than the filter bandwidth. At that point, non-squeezed radiation contributes to $S_f$ limiting the measured squeezing level.
This is a different illustration of the gain-bandwidth trade-off of cavity-based parametric amplifiers~\cite{Clerk:2010dq}. A more detailed discussion of this effect is given in Appendix~\ref{app:filteredSq}.
The three remaining curves show the effect of the Kerr nonlinearity on $S_f$ as a function of the numerically calculated gain [vertical axis of Fig.~\ref{fig:gain}(a)]. As expected, while at low gain all curves overlap, for increasing gain $S_f$ reaches a maximal value, which decreases with Kerr nonlinearity.

Fig.~\ref{fig:gainAndSqueezingNum}(b) compares the squeezing level of a monochromatic current-pumped JPA (up and down triangles), and  a bichromatic current-pumped JPA or monochromatic flux-pumped JPA (circle and squares) for two Kerr nonlinearities. 
As already discussed in Sec.~\ref{sec::gainEta}, the effect of nonidealities on the gain could not be obtained in the case of the monochromatic current-pumped JPA.
To compare pumping schemes on equal footing, squeezing levels are shown as a function of the phase-preserving photon number gain calculated for an ideal DPA using Eq.~\eqref{eq:phasePreservingGain}. Note that the solid curves for the bichromatic current-pumped JPA present the same squeezing levels as in panel~(a), but as a function of the ideal DPA gain.
The curves have similar shapes for both pumping schemes. However, due to the additional cubic corrections, the monochromatic current pump JPA achieves smaller maximal squeezing level than is possible for a bichromatic current pump JPA with single Kerr-type correction. Hence, for the same JPA, going from a monochromatic to a bichromatic current pump allows to significantly increase the maximal squeezing level that can be produced. 
These results are in agreement with what could be expected from intracavity signatures in Sec.~\ref{sec::intracavitySignatures}, where both the second order moments and the Wigner functions presented stronger nonidealities in the monochromatic current pump case.

\begin{figure}[tb]
	\includegraphics[width=\linewidth]{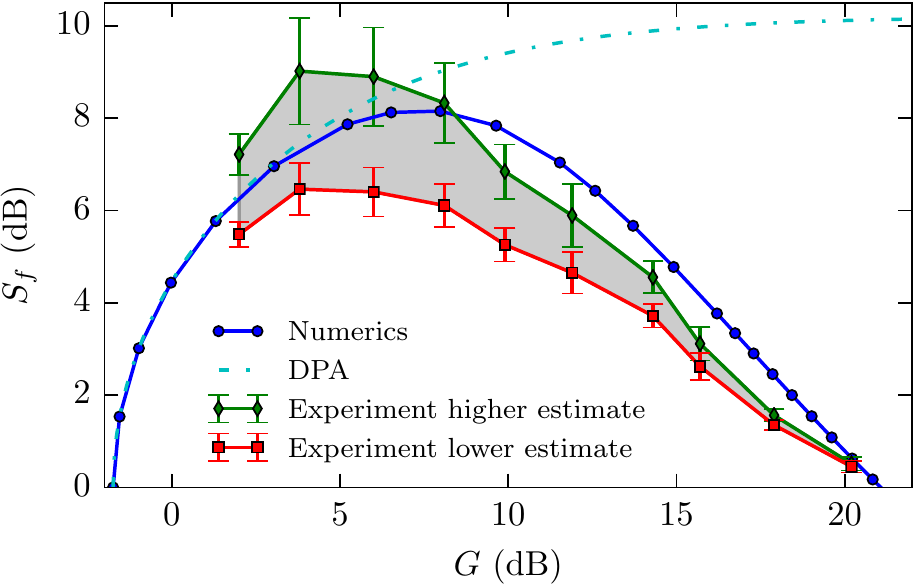}
	\caption{
	Experimental characterization (green triangles and red squares) of the squeezing level of a monochromatic flux-driven JPA with comparison to numerical results (blue circles). Numerical calculations were performed with the parameters $\Lambda/2\pi = -1.55$~MHz, $\kappaTot/2\pi = 130$~MHz and $\gamma= \kappa/10$, obtained from independent measurements (no fitting parameters). {The experimental parameters, setup and method are presented in Appendix~\ref{App::exp}}.
	}
	\label{fig:squeezingExp}
\end{figure}

In order to confirm that these higher-order correction effects explain the experimentally observed reduction in squeezing levels of JPAs~\cite{Castellanos-Beltran:2008vn,Mallet:2011fk,Murch:2013kx,Zhong:2013vn}, we compare our numerical results to experimental data. Using a moment-based reconstruction method, the squeezing level of a flux-pumped JPA is measured~\cite{Silva:2010mi,Eichler_PRL_2011,Menzel_PRL_2012}.
Fig.~\ref{fig:squeezingExp} shows experimental results and numerical data together. The reconstruction technique being highly sensitive to the gain of the measurement chain, the green diamonds (red squares) are the higher (lower) estimate of the squeezing level based on the corresponding measurement chain gain estimate, while the error bars corresponds to statistical error evaluation. The blue circles are numerical results where all parameters of the simulations were obtained from independent experiments with no fitting parameters. As a comparison, the dotted-dashed light-blue line indicates the result expected for an ideal DPA. 
While there is quantitative agreement between numerics and experiment for the maximal squeezing level measured, the overall agreement is only qualitative. 
The small discrepancies could be due to spatial variations in the impedance of the JPA environment that leads to an experimentally observed variation in the gain-bandwidth product of the JPA with increasing gain. 

We note that, contrary to previous hypothesis~\cite{Zhong:2013vn}, our results indicate that the squeezing saturation happens at pump powers below the bifurcation threshold of the JPA. Our results strongly indicate the JPA higher-order correction as the main contributing factor to the experimentally observed decrease of squeezing in the large gain limit.

\begin{figure}[tb]
	\includegraphics[width=\linewidth]{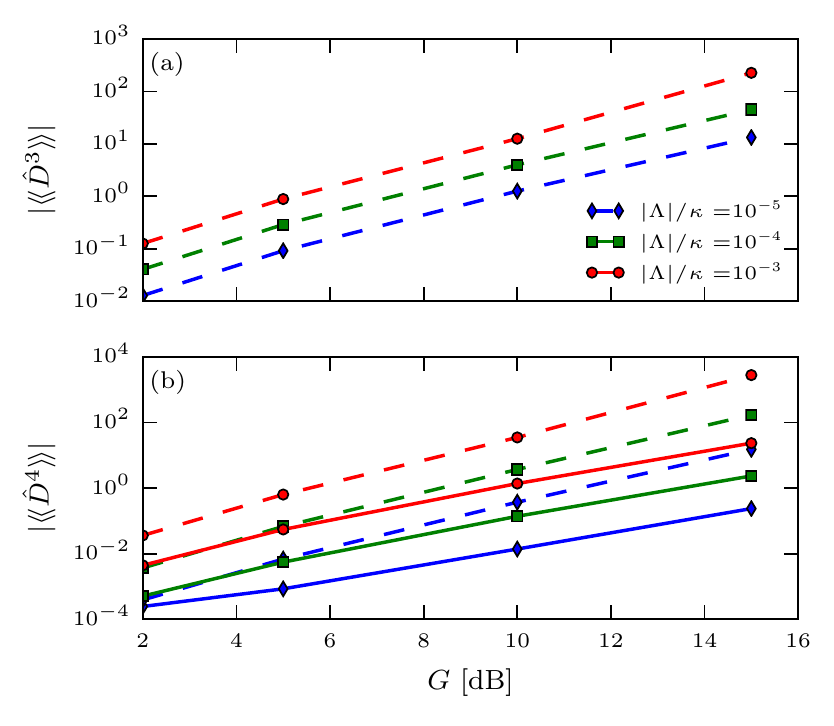}
	\caption{
		Numerical calculation of representative third order (panel (a)) and fourth order (panel (b)) cumulants of the filtered output field. The dashed curves are for the monochromatic current pump (cubic and quartic correction), while the solid curves are for the Kerr quartic correction (bichromatic current pump  or monochromatic flux pump). The third order cumulant is exactly zero in the quartic correction case.
		The filter is a Gaussian filter with 4~MHz bandwidth. In all calculations, we consider $\gamma = \Delta = 0$ and $\kappa/2\pi = 50$~MHz.
	}
	\label{fig:cumulants}
\end{figure}

\subsection{Cumulants}
The Wigner functions of Fig.~\ref{fig:intracavityWigner} clearly indicate that the higher-order corrections lead to non-Gaussian \emph{intracavity} fields. 
In order to characterize the non-gaussianity of the filtered \emph{output} field, we  compute third and fourth order cumulants $\cumulant{\hat D^3}$ and $\cumulant{\hat D^4}$.
In the case of a univariate distribution, the third (fourth) order cumulant can be normalized to define the skew (kurtosis) of the distribution. While such definitions are not as straightforward in the case of the multivariate distribution considered here, the third and fourth order cumulants still characterize the non-Gaussian character of the field.
Indeed, recall that a cumulant of order $n$ is a polynomial of moments of order $n$ and less, and that, for a Gaussian field, only the cumulants of order one and two are non-zero~\cite{Puri:2001kx}.

Fig.~\ref{fig:cumulants}(a) shows numerical calculation of $|\cumulant{\hat D^3}|$ for the JPA with a monochromatic current pump. The cumulant increases following a power law with gain and nonlinearity. The numerical results for the bichromatic current and flux pump are not shown here as they are exactly zero. 
Fig.~\ref{fig:cumulants}(b) shows a fourth order cumulant for both type of corrections. Again, in agreement with the results of Fig.~\ref{fig:intracavityWigner}, non-gaussianity increases with gain and nonlinearity. In addition, we observe that the slope is larger for the monochromatic current pump (cubic corrections).
The other third and fourth order cumulants $\cumulant{\hat D^\dag \hat D^2}$, $\cumulant{\hat D^\dag  \hat D^3}$, and $\cumulant{\hat D^{\dag 2}\hat D^2}$ were also calculated and similar trends were observed (data not shown).

These numerical results present higher-order corrections to the DPA Hamiltonian as a significant source of non-gaussianity in the output field. 
In addition to the experimental data presented below, these corrections could explain previously reported experimental observation of non-Gaussian features for JPAs~\cite{Mallet:2011fk,Zhong:2013vn}.

\subsection{Output field imaging}\label{sec::imaging}
\begin{figure}[tb]
	\includegraphics[width=\linewidth]{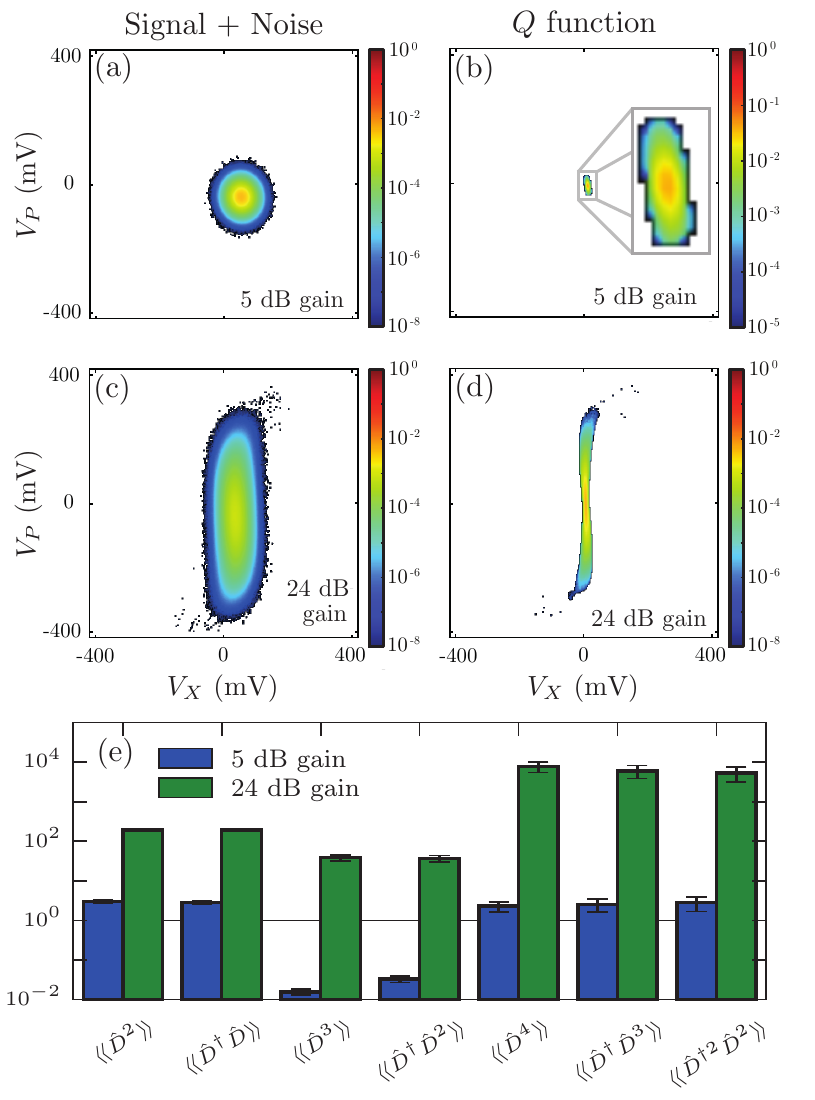}
	\caption{
	Raw phase space images of the JPA measurement chain output (including HEMT noise) as determined via homodyne measurements for (a) 5 dB and (c) 24 dB of phase preserving gain. 
	Each histogram is formed of $256 \times 256$ points where the axes indicate the homodyne quadrature voltages $V_X$, $V_P$ ($\pm 400$~mV range) and the colorscale indicates the relative count intensity for each bin.
	The corresponding JPA output field Q functions inferred via the Lucy-Richardson deconvolution are shown in (b) and (d).
	The deconvolved distributions are plotted on the same measurement voltage range so that the larger areas of the distributions in (a) and (c) reflect the relative contributions of HEMT noise to those measurements. The magnitude of the corresponding output field cumulants, scaled via the gain calibration described in Appendix~\ref{App::exp}, are shown in panel (e).
	\label{fig:qFuncExp}
	}
\end{figure}

In order to further investigate the non-Gaussian features discussed above, we experimentally image the Husimi $Q$-function of the output field of a flux-driven JPA. A deconvolution technique is used to extract the JPA output field from the noisy histograms resulting from sampling the measurement chain output via homodyne detection. This methodology not only provides an experimental analysis of the higher-order cumulants of the output field but also provides a direct image of the \emph{output} field that can be compared with qualitative expectations based on the \emph{intracavity} fields calculated in Fig.~\ref{fig:intracavityWigner}. See Appendix~\ref{App::exp} for experimental details of the method.

Fig.~\ref{fig:qFuncExp} presents a comparison of the output field for low JPA gain (5 dB, top row) and high JPA gain (24 dB, bottom row).
In the figures, both the raw noisy histograms, which are the convolution of the signal with noise due to the amplification chain (left column), and the Husimi $Q$-function extracted through the use of deconvolutions (right column) are shown. 
In addition, panel (e) presents the magnitude of the cumulants up to fourth order extracted from the deconvolved distributions of panels (b) and (d).
As expected from the numerical calculations, the low gain $Q$-function appears Gaussian up to experimental resolution and third and fourth order cumulants are small.
On the contrary, the $Q$-function at high gain is non-Gaussian with a noticeable ``S"-shaped distortion, consistent with expectations based on the \emph{intracavity} fields presented in Figure~\ref{fig:intracavityWigner} and also consistent with a recent semi-classical analysis of the JPA response~\cite{Bienfait:2016yq}. The inferred third and fourth order output field cumulants presented in panel~(e) clearly deviate from the ideal value of zero expected for a Gaussian distribution. However, from the numerical results of Figure~\ref{fig:cumulants} and the symmetry with respect to the centroid of the $Q$-function in panel~(d), one would expect the role of the third order cumulants to be small. To this end, we note that the third order cumulants are smaller than the second order cumulants, in support of the numerical predictions and the imaged $Q$-functions.

To conclude, our experimental results are consistent with the numerically calculated \emph{intracavity} Wigner functions presented in Fig.~\ref{fig:intracavityWigner}. 
While the different orientations of the squeezed ellipse relative to the axes is a consequence of different pump phases, the different orientations relative to the ellipse of the ``S"-like nonideality for the numerically computed \emph{intracavity} field and the measured \emph{output} field is explained by input-output theory.

\section{Conclusion}
\label{sec::conclusion}
In summary, the Kerr-type nonlinear correction is a limiting factor for the measurement quantum efficiency of JPAs and the squeezing level of their output field. This correction also leads to non-Gaussian signatures observed both in the intracavity field Wigner function and the fourth order cumulants of the output field.
Our combined numerical and experimental results allow to explain a broad range of experimental observations, such as the smaller than expected quantum efficiency of the JPA~\cite{Murch:2013uq,Murch:2013kx,Vijay:2012uq,weberThesis},
the saturation and decrease of squeezing at high gain in JPAs~\cite{Zhong:2013vn,Murch:2013kx,Mallet:2011fk,Castellanos-Beltran:2008vn},  and non-Gaussian signatures of the output field~\cite{Zhong:2013vn,Mallet:2011fk}. 
In particular, this work presents a new experimental characterization of the output field of a flux-driven JPA, as well as a direct experimental imaging of the non-Gaussian distortions of the output field.
In addition, we have derived and compared the higher-order corrections to the JPA Hamiltonian for three different pumping schemes. 
Our work shows that, in addition to a Kerr-type quartic correction, cubic terms in the Hamiltonian are important in the case of the monochromatic current pump. These additional corrections lead to larger deviations from the expected DPA behavior such as lower attainable squeezing levels and larger non-Gaussian signatures. 

In short, our results indicate three pathways to improving the performance of the JPA as a squeezer and amplifier. 
First, in the case of a JPA operated with a monochromatic current pump,
moving instead to a bichromatic current pumping scheme eliminates cubic corrections, leading to a greatly increased maximal squeezing level and reduced non-Gaussian signatures. Remarkably, this improvement can be obtained for the same circuit and parameters. Otherwise, at the cost of a small added circuit complexity, but using  only a single drive, flux pumping offers similar advantages. 
Second, our results illustrate clearly the value of designing JPAs with small Kerr nonlinearities. This can be obtained by using SQUID arrays to dilute the nonlinearity~\cite{Castellanos-Beltran:2007ys,Eichler:2013fk} or adding additional linear inductance~\cite{Zhou:2014fk}. Finally, our work emphasizes the phase-sensitivity of the JPA in the high gain regime, and the importance of fully characterizing the phase and frequency dependence of the gain matrix in order to operate at the optimal phases and frequencies where the effects of nonidealities  are minimal.

In complement to the numerical approach considered in this paper, a better understanding of the higher-order corrections discussed might be obtained by considering analytical perturbation theory techniques~\cite{Zhang:2014uq}. Preliminary results are promising~\cite{BoutinMastersThesis}.
Similar analysis for a Josephson junction based traveling-wave parametric amplifier could help the current experimental and theoretical effort~\cite{Yaakobi:2013dz,OBrien:2014zr,White:2015vn,Grimsmo2016,Macklin_Science_2015}. 
Finally, while higher-order corrections hinder the performance of the JPA for squeezing and amplification, it can become a feature for other applications of the JPA such as robust cat state preparation and stabilization~\cite{Puri:2017ee} which can be used for quantum computation~\cite{Ofek:2016} and quantum annealing~\cite{Puri:2017a}.

\acknowledgments{
We thank A. A. Clerk, N. Didier, A. Kamal and M. H. Devoret for fruitful discussions.
This work was supported by the Army Research Office under Grant
W911NF-14-1-0078, FRQ-NT and NSERC. A.W.E. acknowledges support from the US Department of Defense through the NDSEG fellowship program.
Computations were made on the supercomputer Mammouth parallele II from Universit\'{e} de Sherbrooke, managed by Calcul Québec and Compute Canada. 
The operation of this supercomputer is funded by the Canada Foundation for Innovation (CFI), NanoQu\'{e}bec, RMGA and the Fonds de recherche du Québec - Nature et technologies (FRQ-NT). This research was undertaken thanks in part to funding from the Canada First Research Excellence Fund. 
}

\appendix
\section{Solution to the DPA's input-output equations}\label{app::solvingDPA}
For this work to be self-contained, we present the standard solution to the equation of motions of the DPA discussed  in Sec.~\ref{sec::DPA}~\cite{Collett:1984kl,Gardiner:2004fk}. Starting from Eq.~\eqref{eq::EOMDPA1} and introducing the vectors
\begin{equation}
	\hat{\mathbf{a} }
	= \matO{\ah \\ \ah^\dagger},
	\quad
	\hat{\mathbf{a}}_{\mathrm{in}} = 
	\matO{\ain \\ \ain^\dagger} 
	\quad
	\text{and}
	\quad
	\hat{\mathbf{b}}_{\mathrm{in}} = 
	\matO{\bin \\ \bin^\dagger} ,
\end{equation}
the DPA can be described by the system of linear equations
\begin{equation}
	\dot{\hat{\mathbf{a}}}
	= 
	\mathbf{M} \hat{\mathbf{a}}
	+\sqrt{\kappa} \hat{\mathbf{a}}_{\mathrm{in}}
	+\sqrt{\gamma} \hat{\mathbf{b}}_{\mathrm{in}},
\end{equation}
where the matrix $\mathbf{M}$ is
\begin{equation}
	\mathbf{M} =
	\matS{
	-\parO{i \Delta + \frac{\kappaTot}{2} }
	& 
	- i \lambda
	\\
	i \lambda^*
	&
	\parO{i \Delta - \frac{\kappaTot}{2} }
	}.
\end{equation}
This system of linear equations is solved by introducing the Fourier transformed operator
\begin{equation}
	\overline{a}\parS{\omega}
	= 
	\int_{-\infty}^{\infty}
	\drm t \,  \erm^{i \omega t} \ah(t).
\end{equation}
We note that $\overline{a^\dagger}[\omega]$ the Fourier transform of the creation operator is related to the adjoint of $\overline{a}[\omega]$ by $\overline{a^\dagger}[\omega] = \overline{a}^\dagger[-\omega]$. Thus, the solution to the linearized equations of motions in Fourier space is~\cite{Collett:1984kl,Gardiner:2004fk}
\begin{equation}
	\overline{\mathbf{a}}\parS{\omega} = 
	-\parO{\mathbf{M} + i \omega \mathbf{1}}^{-1}
	\parO{
		\sqrt{\kappa}
		\overline{\mathbf{a}_{\mathrm{in}}}\parS{\omega}
		+
		\sqrt{\gamma}
		\overline{\mathbf{b}_{\mathrm{in}}}\parS{\omega}
	}.
\end{equation}
Using the input-output boundary condition Eq.~\eqref{eq:boundary}, this result leads to Eq.~\eqref{eq::DPA_inputOutput} of the main text.

For completeness, we obtain from this expression the general expressions for the second order centered moments of the intracavity field used in Sec.~\ref{sec::intracavitySignatures}
\begin{align}
	N_{\mathrm{DPA}} &= 
	\frac{\abs{\lambda}^2}{2 \parO{\Delta^2 + \kappaTot^2/4- \abs{\lambda}^2}},
	\\
	M_{\mathrm{DPA}} &=
	\frac{- \lambda \parO{\Delta + i \kappaTot/2}}{2 \parO{\Delta^2 + \kappaTot^2/4- \abs{\lambda}^2}}
	.
\end{align}
Using the input-output boundary condition Eq.~\eqref{eq:boundary}, the output field spectrums are~\cite{Collett:1984kl}
\begin{align}
 	\overline{N}_{\mathrm{out}}[\omega] 
 	&=
 	\frac{\abs{\lambda}^2 \kappa \kappaTot}{ \parO{\Delta^2 + \kappaTot^2/4-\omega^2-\abs{\lambda}^2 }^2 + \kappaTot^2 \omega^2}
 	\label{eq:N_out}
 	\\
 	\overline{M}_{\mathrm{out}}[\omega] 
 	&=
 	\frac{
 		-i \kappa \lambda \parS{\parO{\kappaTot-i \Delta}^2+\omega^2+ \abs{\lambda}^2}
 	}{ \parO{\Delta^2 + \kappaTot^2/4-\omega^2-\abs{\lambda}^2 }^2 + \kappaTot^2 \omega^2}
 	,
 	\label{eq:M_out}
\end{align} 
with $\overline{N}_{\mathrm{out}}[\omega]  = \int \drm \overline{\omega}  \langle  
\overline{\aout^\dag}[\omega] \aout [\omega]
\rangle/2 \pi$, and similarly 
$\overline{M}_{\mathrm{out}}[\omega]  = \int \drm \overline{\omega}  \langle  
\overline{\aout}[\omega] \aout [\omega]
\rangle/2 \pi$. These expressions allows to calculate analytically the DPA filtered output field spectrums presented in Sec.~\ref{sec::filteredOutput}.

\section{Details of the displacement transformations}\label{app::doubleDisplacement}
In this appendix, we detail the displacement transformations used to derive 
the Hamiltonian Eq.~\eqref{eq::hSinglePump} for the monochromatic current-pumped JPA and the Hamiltonian Eq.~\eqref{eq::bichromaticPumpHamiltonianAfterDisplacement} for the bichromatic current-pumped JPA.
The unitary displacement transformation is~\cite{Walls:2008fk}
\begin{equation}
\hat D (\beta) = \exp \parO{ \beta \ah^\dag - \beta^* \ah},
\end{equation}
with $\hat D^\dag(\beta) \ah \hat D(\beta) = \ah +\beta$, $\hat D(\beta_1) \hat D(\beta_2) = \hat D(\beta_1+ \beta_2)$, and $\beta, \beta_{(1,2)}$ scalar complex numbers.

Applying this transformation on the driven Kerr Hamiltonian
\begin{equation}
	\hat H_i = \Delta \ah^\dag \ah + \Lambda \ah^{\dag 2} \ah ^2 + \epsilon(t) \ah ^\dag  + \epsilon^*(t) \ah,
\end{equation} 
with $\Delta= \tilde{\omega}_0 - \omega_{\mathrm{rot}}$ the detuning between the cavity and the rotating frame frequency,
one obtains
\begin{equation}
	\hat H_i' = \hat D^\dag (\beta) \hat H \hat D(\beta) - i \dot{\hat D}(\beta) \hat D^\dag (\beta).
\end{equation}
Dropping constant terms and taking $\ah \to \dhat$ to emphasize the frame change, one obtains the Hamiltonian
\begin{equation}
	\begin{split}
	\hat H_i' &= \parO{ \Delta+ 4 \Lambda \abs{\beta}^2  } \dhat^\dag \dhat
	\\ &
		+ \Lambda \parS{ \beta^2 \dhat^{\dag 2} + \beta \dhat^{\dag 2} \dhat   + h.c.} + \Lambda \dhat^{\dag 2} \dhat^2,
	\end{split}
\end{equation}
where the linear pump term has been canceled by choosing the displacement parameter $\beta$ such that
\begin{equation}
	i \dot \beta = \epsilon(t) + \parO{ \Delta + 2 \Lambda \abs{\beta}^2
	 - i \frac{ \kappaTot }{2}  } \beta.
	 \label{eq:betaDot}
\end{equation}
The additional term $-i \kappaTot \beta /2$ originates from applying the displacement transformation on the master equation Eq.~\eqref{eq:ME} instead of only the Hamiltonian.

In the single current-pump case of Sec.~\ref{subsec::singlePump}, in a frame rotating at the pump frequency ($\omega_{\mathrm{rot}} = \wpump$) the pump is simply $\epsilon(t) = \epsilon$, and one obtains the results of the main text by taking $\alpha = \beta$.

In the double pump case of Sec.~\ref{subsec::doublePump}, in a frame rotating at the average pump frequency $\Omega_{12} = (\omega_1 + \omega_2)/2 $,
the driving term is 
\begin{equation}
\epsilon(t) = \epsilon_1\, \erm^{+i \Delta_{12} t/2}
		 	+
		 	\epsilon_2\, \erm^{-i \Delta_{12} t/2},
		 	\label{eq:epsTwoPumps}
\end{equation}
with the detuning between the pumps $\Delta_{12} = \omega_1 - \omega_2$.
Inserting Eq.~\eqref{eq:epsTwoPumps} in Eq.~\eqref{eq:betaDot} and taking the ansatz $\beta = \alpha_1 \erm^{+i \Delta_{12}t /2} + \alpha_2 \erm^{-i \Delta_{12}t /2} $ leads to the coupled nonlinear differential equations

\begin{align}
\begin{split}
	i\dot {\alpha}_1
	&=
	\epsilon_1 
	+
	\parO{ 
	\tilde{\omega}_0 - \omega_1
	+ 2 \Lambda \abs{\alpha_1  }^2
	-i \frac{\kappaTot}{2}
	}\alpha_1
	\\
	&\,+ 2 \Lambda \parO{
		2\abs{\alpha_2 }^2
		+
		\alpha_1 \alpha_2^* \erm^{-i \Delta_{12}t}
	}\alpha_1,
\end{split}
\\
\begin{split}
	i\dot {\alpha}_2
	&=
	\epsilon_2
	+
	\parO{ 
	\tilde{\omega}_0 - \omega_2
	+ 2 \Lambda \abs{\alpha_2  }^2
	-i \frac{\kappaTot}{2}
	}\alpha_2
	\\
	&\,+ 2 \Lambda \parO{
		2\abs{\alpha_1 }^2
		+
		\alpha_2 \alpha_1^* \erm^{i \Delta_{12}t}
	}\alpha_2,
\end{split}
\end{align}
and to the Hamiltonian of Eq.~\eqref{eq::bichromaticPumpHamiltonianAfterDisplacement}.
For $\Delta_{12}  \gg 2 \abs{ \Lambda \alpha_1 \alpha_2}$, we can neglect the rotating terms under the RWA. 

Since the parametric resonance condition only depends on the sum of the pump frequencies, and the amplitude of the classical field is bounded by the parametric threshold, one can choose $\Delta_{12}$ in order to enforce the validity of the RWA. 
Under this approximation, the equations are the same as Eq.~\eqref{eq::alphaDot} in the monochromatic current pump case except that the equations are coupled through the frequency shift induced by the cavity population at both pump frequencies.


\section{Expansion of the flux-modulated Josephson energy}\label{app::fluxPumpCoeff}
In this appendix, we complement Sec.~\ref{subsec::fluxPump} by giving the analytical expressions for the Fourier coefficients of the Josephson energy of a flux modulated SQUID used in Eq.~\eqref{eq::fourierExpandFluxPump}.
Using the Jacobi-Anger formula~\cite{Arfken:2005fk}
\begin{equation}
	\exp\parO{
	 i x \cos \theta
	}
	= J_0(x) + 2 \sum_{n=0}^{\infty} i^n J_n(x) \cos n \theta,
\end{equation}
where $J_n(x)$ is the $n^{\mathrm{th}}$ Bessel function of the first kind, one obtains the Fourier coefficients
\begin{align}
	E_J^{(0)} &= E_J J_0(\df) \cos F
	,\\
	E_J^{(2n-1)} &= 2 E_J (-1)^n  J_{2n-1}(\df) \sin F
	,\\
	E_J^{(2n)} &= 2E_J (-1)^n  J_{2n}(\df) \cos F
	,
\end{align}
with $n \in \parC{1,2,3 \dots}$.
	
In the case of a small amplitude flux pump ($\df \ll 1$), the Bessel functions can be expanded. The leading term of each coefficient is such that
\begin{equation}
	E_J^{(n)} \propto \frac{1}{n!} \parO{\frac{\df}{2}}^n.
\end{equation}
 More explicitly, the first three coefficients of the Fourier expansion are, to leading order in $\df$, 
\begin{align}
	E_J^{(0)} &\approx  E_J \cos F
	,\\
	E_J^{(1)} &\approx - E_J \df \sin F
	,\\
	E_J^{(2)} &\approx  - \frac{E_J \df^2 \cos F}{4}
	,
\end{align}
in agreement with the expressions of Refs.~\cite{Wustmann:2013uq,Krantz:2013vn}.

\section{Experimental setup and methods}\label{App::exp}
This appendix presents details of the experimental setup and techniques used to obtain the experimental data for a flux-driven JPA presented in Figs.~\ref{fig:squeezingExp} and \ref{fig:qFuncExp}. 

\subsection{Experimental setup}
We focus our characterization on an aluminum, lumped-element JPA.  This design is of particular interest as it has been widely adopted for superconducting qubit readout~\cite{Vijay:2011ve,Mutus:2013yu,Murch:2013uq}.  The device consists of a capacitance (3.2 pF) shunted by a SQUID ($L_\mathrm{J}\left(\Phi=0\right)$ = 45 pH).  From simulations we estimate the geometric inductance of our design is 35 pH leading to a participation ratio $p=0.8$. From these we estimate a Kerr nonlinearity $\Lambda/2\pi= -1.55$~MHz and from gain-bandwidth measurements a cavity damping rate $\kappaTot/2 \pi = 130$~MHz. 
This empirical value of $\kappa$ is lower than expected from the lumped-element capacitance; we attribute this difference to spatial impedance variations in the JPA environment~\cite{Mutus:2014ve}.
The SQUID is flux-pumped to provide up to 25 dB of gain using an on-chip bias line~\cite{Yamamoto:2008dp} and the device is mounted with a 180$^{\circ}$ hybrid launch to reject common mode noise. 
The amplifier is shielded by an aluminum box and mounted at the base plate of a dilution refrigerator ($\sim$ 20 mK).  The signal from the JPA is further amplified with a commercial HEMT amplifier at 4K and subsequent room temperature amplifiers before being downconverted and digitized in a homodyne measurement. 

The homodyne measurement samples the values of the conjugate quadrature components $\hat{X}$ and $\hat{P}$ described by the complex quadrature operator $\hat{S} = \hat{X}+i\hat{P}$~\cite{Silva:2010mi}. For each measurement we averaged 256 consecutive voltage samples acquired at 1 GS/s, effectively filtering $\hat{S}$ so as to analyze only highly-squeezed spectral components near the cavity frequency. The flux pump at frequency 2$\omega$ is generated using a frequency doubler and the same microwave source that produces the local oscillator tone for the homodyne setup, allowing for good phase stability over the course of extended measurements. In addition, the JPA is hooked up to a switch on the base stage of the dilution refrigerator. The other port of the switch connects to a qubit dispersively coupled to a tunable resonator~\cite{Ong_PRL_2011}.  Through measurements of the qubit Stark shift, this setup enables a precise calibration of the power gain between the JPA and the ADC~\cite{Macklin_Science_2015}, a necessary input for our analysis.

\subsection{Single-path reconstruction method}\label{app:expSPM}
To characterize the properties of the output field we utilize the single path reconstruction method developed by Eichler~\textit{et al.}~\cite{Eichler_PRL_2011}. We construct two histograms of $\hat{S}$, corresponding to the distribution with the JPA pump on, $D^{[\rho]}\left(S\right)$, and to the distribution with the JPA pump off, $D^{[\left|0\rangle\langle0\right|]}\left(S\right)$. We interleave the two measurements to mitigate the effects of experimental drift. We analyze the moments of the two histograms, $\langle ( \hat{S}^\dagger)^n \hat{S}^m \rangle_{\rho}$ and $\langle ( \hat{S}^\dagger)^n \hat{S}^m \rangle_{\left|0\rangle\langle0\right|}$, to infer the normally-ordered moments of the output field, $\langle (\hat a^\dagger )^n \hat a^m \rangle$, through the expressions:  
\begin{multline}\label{signalplusnoise}
\langle ( \hat{S}^\dagger)^n \hat{S}^m \rangle_{\rho} = G_\mathrm{c}^{(n+m)/2} \sum \limits_{i,j=0}^{n,m} {m \choose j} {n \choose i}\langle (\hat a^\dagger )^i \hat a^j \rangle \\ \times \langle \hat h^{n-i} (\hat h^\dagger)^{m-j} \rangle
\end{multline}  
and 
\begin{equation}\label{vacuumplusnoise}
\langle ( \hat{S}^\dagger)^n \hat{S}^m \rangle_{\left|0\rangle\langle0\right|} = G_\mathrm{c}^{(n+m)/2} \langle \hat h^{n} ( \hat h^\dagger)^{m} \rangle.
\end{equation}
Here $G_\mathrm{c}$ is the gain of the measurement chain between the JPA output and the ADC in the photon number basis~\cite{Menzel_PRL_2012, Menzel_thesis} and $\hat h$ is an effective noise mode dominated by the added noise of the HEMT amplifier.  
Experimentally, the power gain of the measurement chain was measured to be between 100.1~dB and 100.5~dB. These bounds on the power gain are reflected in the systematic uncertainty in the squeezing levels presented in Figure~\ref{fig:squeezingExp}.
Note that Eq.~\ref{vacuumplusnoise} is a limiting case of Eq.~\ref{signalplusnoise} when the JPA pump is off and $\hat a$ is in the vacuum state.  We iteratively solve these equations to compute the moments of the JPA output field and also quantify the cumulants of the output field, denoted by $\langle\! \langle (\hat a^\dagger )^n \hat a^m \rangle\!\rangle$. For Gaussian states, such as ideal squeezed states, the cumulants are zero for $n + m > 2$.  At each gain setting we histogram $10^8$ noise measurements which provides sufficient resolution to infer the field moments up to fourth order.

\subsection{Imaging the output field using deconvolutions}
To directly image the distortions of the output field due to nonidealities, we implement a complementary analysis technique based on applying series of deconvolutions to $D^{[\rho]}\left(S\right)$ and $D^{[\left|0\rangle\langle0\right|]}\left(S\right)$. This method relies on the fact that these discrete distributions can be approximated as a convolution of quasi-probability distributions for the JPA output field and an effective noise mode of the measurement chain~\cite{Kim_PRA_1997}:
\begin{equation}\label{LRone}
D^{[\rho]}\left(S\right) \approx \frac{1}{G_\mathrm{c}}(Q_{\mathrm{JPA}} 
\star P_{\mathrm{HEMT}})(\frac{\alpha}{\sqrt{G_\mathrm{c}}})
\end{equation}
and
\begin{equation}\label{LRtwo}
D^{[\left|0\rangle\langle0\right|]}\left(S\right) \approx \frac{1}{G_\mathrm{c}}(Q_{\mathrm{vacuum}} \star P_{\mathrm{HEMT}})(\frac{\alpha}{\sqrt{G_\mathrm{c}}}).
\end{equation}
Here $Q_{\mathrm{JPA}}$, $Q_{\mathrm{vacuum}}$, and $P_{\mathrm{HEMT}}$ are Husimi $Q$ and Glauber-Sudarshan $P$ representations that describe the JPA output field, the vacuum state, and the noise mode~\cite{Eichler_PRA_2012}. Given that the $Q$ function can be obtained from the Wigner function via a Gaussian smoothing filter, we expect the distortions imaged by this deconvolution technique to be characteristic of both phase-space representations.

To reconstruct the output field, we first deconvolve $D^{[\left|0\rangle\langle0\right|]}\left(S\right)$ with $Q_{\mathrm{vacuum}}$ to obtain $\frac{1}{G_\mathrm{c}}P_{\mathrm{HEMT}}$ and then deconvolve $D^{[\rho]}\left(S\right)$ with $\frac{1}{G_\mathrm{c}}P_{\mathrm{HEMT}}$ to obtain $Q_{\mathrm{JPA}}$. Both deconvolutions are performed in Matlab using the Lucy-Richardson method~\cite{Biggs_97}.
For the measurements in Figure~\ref{fig:qFuncExp}, the power gain of the measurement chain was increased by 5~dB relative to the bounds quoted in Appendix~\ref{app:expSPM}. We empirically found this larger gain led to improved resolution of the low-density tails of the output field Q functions. In all cases, inferences of the output field cumulants obtained via the moment-based reconstruction agreed with inferences of the cumulants calculated from the Q functions.

\section{Numerical calculation of filtered moments}	\label{app::filteredMoments}
In Sec.~\ref{sec::filteredOutput}, we calculate moments of the filtered output field. Here, we give the details of the numerical technique used to obtain the results presented there. As an illustration of the technique, we consider the third order moment
\begin{equation}
	\begin{split}
	\langle \hat D^\dag \hat D^2 \rangle  &= 
	\iiint_{-\infty}^{\infty} \drm t_1 \drm t_2 \drm t_3 \, f(-t_1)f(-t_2)
	\\ &\qquad\times
	f(-t_3) 
		M_{t_1,t_2,t_3} ,
	\end{split}
\end{equation}
with the three-times correlation function
\begin{align}
\begin{split}
	M_{t_1,t_2,t_3} &=\langle \aout^\dag(t_1)\aout(t_2)\aout(t_3) \rangle
	\\
	&=
	\kappa^{3/2}\langle \ah^\dag(t_1)\ah(t_2)\ah(t_3) \rangle,
\end{split}
\end{align}
where the last equality is valid for a vacuum input field~\cite{Gardiner:1985ly}.

In order to use the quantum regression formula, we separate the integral in a sum over all possible time-orderings
\begin{align}
	\begin{split}
	\langle\hat D^\dag \hat D^2 \rangle 
	&= 
	\iint_{0}^{\infty} \drm t_1 \drm t_2  \, F_{t_1,t_2}
	\parS{
		M_{t_1+t_2,t_1,0}
		\right.\\ & \quad\left.
		+
		M_{t_1,t_1+t_2,0}
		+
		M_{0,t_1+t_2,t_1}
	},
	\end{split}
\end{align}
where, using the invariance under time-translation for a steady-state and assuming a purely real filter, we have defined
\begin{equation}
	F_{t_1,t_2} = \int_{-\infty}^{\infty} \drm t_3 \, f(-t_1-t_2-t_3) f(-t_2-t_3) f(-t_3).
\end{equation}
Finally, we integrate over each correlation function that we calculate using the general formula of the quantum regression result~\cite{Gardiner:2004fk}. For the three-time correlation function, we obtain
\begin{align}
	\begin{split}
	M_{\tau,t_1,0} &= 
	\mathrm{Tr}\parC{\aout^\dag V_{\tau,t_1}\parS{\aout V_{t_1,0} \parO{ \aout \hat \rho_{\mathrm{ss}}}}	} 
	\\
	M_{t_1,\tau,0} &= 
	\mathrm{Tr}\parC{\aout V_{\tau,t_1} [V_{t_1,0}\parO{ \aout \hat \rho_{\mathrm{ss}}}\aout^\dag ]} 
	\\
	M_{0,\tau,t_1} &= 
	\mathrm{Tr}\parC{\aout V_{\tau,t_1} [\aout V_{t_1,0}  (\hat \rho_{\mathrm{ss}}\aout^\dag) ] },
	\end{split}
\end{align}
with $V(t_1,0)$ the evolution superoperator from $t=0$ to $t=t_1$, and $\tau = t_1+t_2$. Numerically, the evolution is performed by integrating the master equation using the Runge-Kutta solver of the GSL numerical library~\cite{Galassi:2011uq}. 

\section{Squeezing level of the filtered DPA output field}
\label{app:filteredSq}

In addition to the numerical approach of Appendix~\ref{app::filteredMoments}, the results of Figs.~\ref{fig:gainAndSqueezingNum} and~\ref{fig:squeezingExp} for the DPA can be obtained more simply by integrating the analytical expressions of Eqs.~\eqref{eq:N_out} and~\eqref{eq:M_out} using 
\begin{align}
	\langle \hat D^\dag  \hat D \rangle  = \int_{-\infty}^\infty\!\! \drm \omega \, \abs{\overline{f}[\omega]}^2 \overline{N}_{\mathrm{out}}[\omega] = N_f,
	\\
	 \langle \hat D^2 \rangle   = \int_{-\infty}^\infty\!\! \drm \omega  \abs{\overline{f}[\omega]}^2 \overline{M}_{\mathrm{out}}[\omega] = M_f,
\end{align}
with $\overline{f}[\omega]$ the Fourier transform of a real time-domain filter function $f(t)$.
In particular, in the case of Fig.~\ref{fig:gainAndSqueezingNum} where $\kappaTot = \kappa$ and $\Delta=0$, the minimal variance of the filtered output field is
\begin{equation}
	 \langle \Delta \hat X^2_{\mathrm{Min} }  \rangle 
	 = 
	\frac{ 1 }{2}
	\int
	\drm \omega \, \abs{\overline{f}[\omega]}^2
	 	  \parO{ 1 - \frac{ 2 \kappa \abs{\lambda} }{ \parO{ \kappa/2 + \abs{\lambda}  }^2 + \omega^2} }.
\end{equation}
As mentioned in the main text, the integrand is minimal (maximal squeezing) at the center frequency $\omega=0$, and will increase (reduced squeezing) away from this frequency. Thus, as discussed in Sec.~\ref{subsec:squeezingLevel}, the minimal variance of the filtered field will always be equal, or larger than, the variance at the center frequency. This limits the squeezing level of the JPA filtered output field, even without any nonidealities. 

To further illustrate this effect, we consider the case of a narrow-band filter centered at frequency $\omega_0$ such that $\abs{\overline{f}[\omega]}^2 \approx \delta(\omega-\omega_0)$. In that case, the squeezing level is 
\begin{equation}
	S_f(\omega_0) = 1+ \frac{ 2 \kappa \abs{\lambda} }{ \parO{ \kappa/2 - \abs{\lambda}  }^2 +\omega_0^2},
	\label{eq:sfw0}
\end{equation}
which only diverges for $\omega_0=0$, and otherwise reaches a finite maximal value of $1+ \kappa^2/\omega_0^2$ even at the parametric threshold. Neglecting the constant background, the half-width at half maximum of $S_f(\omega_0)$ is $\kappa/2-\abs{\lambda}$, which will decrease with increasing gain reaching zero at the parametric threshold. This is a different illustration of the gain-bandwidth trade-off of cavity-based parametric amplifiers~\cite{Clerk:2010dq}.

\bibliographystyle{apsrev4-1}
%

\end{document}